\documentclass[bibtex]{aa}  

\usepackage{graphicx}
\usepackage{caption}
\usepackage{txfonts}
\usepackage{natbib}
\DeclareUnicodeCharacter{00A0}{~}
\usepackage[flushleft]{threeparttable}
\usepackage{color}
\begin{document}

   \title{Chemical content of the circumstellar envelope of\\ the oxygen-rich AGB star R Dor}
   \subtitle{Non-LTE abundance analysis of CO, SiO, and HCN}
    \titlerunning{Chemical content of the circumstellar envelope of\\ the oxygen-rich AGB star R Dor}
   \author{M. Van de Sande \inst{\ref{inst1}} \and
           L. Decin \inst{\ref{inst1}} \and
           R. Lombaert \inst{\ref{inst2}} \and
           T. Khouri \inst{\ref{inst2}} \and
           A. de Koter \inst{\ref{inst3}, \ref{inst1}} \and
           F. Wyrowski \inst{\ref{inst4}} \and
           R. De Nutte \inst{\ref{inst1}} \and
           W. Homan \inst{\ref{inst1}}
          }

   \institute{Department of Physics and Astronomy, Institute of Astronomy, KU Leuven, Celestijnenlaan 200D, 3001 Leuven, Belgium \label{inst1} \\ \email{marie.vandesande@kuleuven.be}
         \and Department of Earth and Space Sciences, Chalmers University of Technology, Onsala Space Observatory, 439 92 Onsala, Sweden  \label{inst2} 
         \and Astronomical Institute Anton Pannekoek, University of Amsterdam, Science Park 904, PO Box 94249, 1090 GE Amsterdam, The Netherlands \label{inst3}
         \and Max-Planck-Institut f\"ur Radioastronomie, Auf dem H\"ugel 69, 53121 Bonn, Germany  \label{inst4}
             }
             
   \date{Received June 2, 2017; accepted }

%-------------------------------------------------------------------
\abstract
% context heading (optional)
{The stellar outflows of low- to intermediate-mass stars are characterised by a rich chemistry. Condensation of molecular gas species into dust grains is a key component in a chain of physical processes that leads to the onset of a stellar wind. 
In order to improve our understanding of the coupling between the micro-scale chemistry and macro-scale dynamics, we need to retrieve the abundance of molecules throughout the outflow.
}
% aims heading (mandatory)
{Our aim is to determine the radial abundance profile of SiO and HCN throughout the stellar outflow of R Dor, an oxygen-rich AGB star with a low mass-loss rate. 
SiO is thought to play an essential role in the dust-formation process of oxygen-rich AGB stars. The presence of HCN in an oxygen-rich environment is thought to be due to non-equilibrium chemistry in the inner wind.
}
% methods heading (mandatory)
{We have analysed molecular transitions of CO, SiO, and HCN measured with the APEX telescope and all three instruments on the \emph{Herschel Space Observatory}, together with data available in the literature. Photometric data and the infrared spectrum measured by ISO-SWS were used to constrain the dust component of the outflow.
Using both continuum and line radiative transfer methods, a physical envelope model of both gas and dust was established.
We have performed an analysis of the SiO and HCN molecular transitions in order to calculate their abundances.
}
% results heading (mandatory)
{We have obtained an envelope model that describes the dust and the gas in the outflow, and determined the abundance of SiO and HCN throughout the region of the stellar outflow probed by our molecular data. 
For SiO, we find that the initial abundance lies between $5.5 \times 10^{-5}$ and $6.0 \times 10^{-5}$ with respect to H$_2$. The abundance profile is constant up to $60\ \pm 10\ R_*$, after which it declines following a Gaussian profile with an $e$-folding radius of $3.5 \pm 0.5 \times 10^{13}$ cm or $1.4 \pm 0.2\ R_* $.
For HCN, we find an initial abundance of $5.0 \times 10^{-7}$ with respect to H$_2$. The Gaussian profile that describes the decline starts at the stellar surface and has an $e$-folding radius $r_e$ of $1.85 \pm 0.05 \times 10^{15}$ cm or $74 \pm 2 \ R_*$.
}
% conclusions heading (optional)
{We cannot to unambiguously identify the mechanism by which SiO is destroyed at $60\ \pm 10\ R_*$. 
The initial abundances found are larger than previously determined (except for one previous study on SiO), which might be due to the inclusion of higher-$J$ transitions.
The difference in abundance for SiO and HCN compared to high mass-loss rate Mira star IK Tau might be due to different pulsation characteristics of the central star and/or a difference in dust condensation physics. 
}


   \maketitle

%%%%%%%%%%%%%%%%%%%%%%%%%%%%%%%%%%%%%%%%%%%%%%%%%%%%%%%
\section{Introduction}
%%%%%%%%%%%%%%%%%%%%%%%%%%%%%%%%%%%%%%%%%%%%%%%%%%%%%%%

During the asymptotic giant branch (AGB) phase, stars with an initial mass up to 8 M$_\odot$ undergo strong mass loss via a stellar wind. This removes the outer stellar layers and contributes significant amounts of gas and dust to the interstellar medium.
The outflow of material creates a circumstellar envelope (CSE), which is characterised by a rich chemistry. The type of chemistry is characterised by the elemental carbon-to-oxygen abundance ratio: oxygen-rich M-type stars have C/O < 1, carbon-rich stars have C/O > 1, and S-type stars have C/O $\approx$ 1.

The driving mechanism of the AGB outflow is thought to be a two stage process: stellar pulsations initiate dust formation near the stellar surface by lifting material to cooler regions, radiation pressure onto the dust grains then drives the stellar wind.
This scenario works well for carbon stars, as the refractory carbonaceous particles can exist relatively close to the stellar surface and are relatively opaque to the stellar photons \citep{Mattsson2010}.
However, it does not reproduce the observed mass-loss rates for M-type stars. 
Opaque oxygen-based dust species cannot exist sufficiently close to the star to provide the driving needed to account for the observed mass-loss rate \citep{Woitke2006}. 
Several solutions have been suggested, such as the formation of both carbon and silicate grains and micron-sized Fe-free silicate grains \citep{Hofner2007} and scattering on large ($\sim$ 300 nm), translucent grains \citep{Hofner2008}. 

In this paper, we investigate the stellar wind of the M-type AGB star R Doradus, henceforth \object{R Dor}.
R Dor is classified as a semiregular variable of subtype b (SRb) and switches between pulsation modes with periods of 332\,d and 175\,d \citep{Bedding1998}. It is characterised by a low mass-loss rate of $\sim$ 1 $\times$ 10$^{-7}$ M$_\odot$ yr$^{-1}$.
%$\sim$ 9 $\times$ 10$^{-8}$ M$_\odot$ yr$^{-1}$. 
Thanks to its large angular diameter of 57 $\pm$ 5 mas, the largest on the sky after the Sun, the star has been extensively observed, also with all instruments on the \emph{Herschel Space Observatory}. As a result, a multitude of molecules has been detected in its CSE.
\citet{Norris2012} have observed a close halo of large translucent dust grains around the star, which corroborates the proposed solution for the wind driving mechanism by \citet{Hofner2008}. Recently, \citet{Khouri2016} have also observed a close halo of dust grains using the SPHERE/ZIMPOL instrument on the \textit{Very Large Telescope}.

In order to better comprehend the physics underlying the wind driving mechanism, it is necessary to understand both the dynamical and chemical structure of the wind. More specifically, we need to understand the coupling between the small-scale chemistry occurring within the stellar wind and its large-scale dynamics, since the chemical processes at play in the innermost regions of the wind are responsible for dust formation. The chemical pathways by which the first dust particles condense are still unknown, as is their further growth.
By observing molecular line emission from the CSE, its radial abundance profile can be constrained. These profiles contain information on the ongoing circumstellar chemistry of a given molecular species, e.g. depletion of the species onto dust grains, and may be confronted with forward chemistry models that predict the abundance stratification of molecules throughout the stellar wind (see, e.g., \citeauthor{Gobrecht2016}~\citeyear{Gobrecht2016}, \citeauthor{Li2016}~\citeyear{Li2016}). 
Such models require certain stellar and dynamical parameters together with a set of parent species and their initial abundances as input, and solve a chemical network. 
For M-type AGB stars, the chemistry occuring within the stellar wind has been studied for IK Tau, an M-type AGB star with a high mass-loss rate of $\sim$ 8 $\times$ 10$^{-6}$ $M_\odot$ yr$^{-1}$ \citep{Decin2010}.
Gas-phase chemistry and dust formation within the inner wind has been studied by \citet{Duari1999}, \citet{Cherchneff2006} and \citet{Gobrecht2016}. The outer wind chemistry has been studied by \citet{Willacy1997} and \citet{Li2016}.
The intermediate wind has, so far, not been studied in great detail. 

Here we investigate the dynamical structure of the stellar wind of the low mass-loss rate AGB star R Dor ($\dot{\mathrm{M}} = 1.25 \times 10^{-7}$ M$_\odot$ yr$^{-1}$) and the abundance profiles of SiO and HCN throughout the stellar wind. 
The retrieved abundance profiles may place firm constraints on future forward chemistry modelling of the (outer) wind. Moreover, they are complementary to those retrieved by \citet{Decin2010} for IK Tau, and allow us to compare the effect of the wind dynamics on the chemistry for different mass-loss rates.
SiO is one of the most abundant molecules in the CSEs of O-rich AGB stars. It is abundantly present in the dust nucleation zone \citep{Gail1998, Duari1999}.
It is thought to play an important role during dust condensation, either via surface addition processes to small titanium oxide or aluminium oxide clusters or via nucleation, with dimerization as initial condensation process \citep{Goumans2012}. 
The presence of HCN in the CSEs of M-type AGB stars is thought to point to non-thermal equilibrium chemistry induced by shocks in the  atmosphere of the central star occuring in the inner wind \citep{Duari1999,Cherchneff2006}, making it a good probe for understanding the chemical processes at play. As HCN does not participate in the formation of dust grains and is chemically quite stable, it is expected to travel unaltered through the entire CSE until its photodissociation radius \citep{Duari1999}.

The observational data used in this paper are presented in Sect. \ref{sect:Data}. 
Sect. \ref{sect:Methodology} describes our modelling methodology. 
In Sect. \ref{sect:Results} we present our results, which are discussed in Sect. \ref{sect:Discussion}. We end with the conclusions in Sect. \ref{sect:Conclusions}.

%%%%%%%%%%%%%%%%%%%%%%%%%%%%%%%%%%%%%%%%%%%%%%%%%%%%%%%
\section{Observational data}			\label{sect:Data}
%%%%%%%%%%%%%%%%%%%%%%%%%%%%%%%%%%%%%%%%%%%%%%%%%%%%%%%

The molecular data used were partly obtained from our own observing program scheduled at APEX. 
R Dor has also been observed with all three instruments onboard of the \emph{Herschel Space Observatory}.  
Additional data has been obtained from the literature.
All resolved transitions used in this paper are listed in Table \ref{table:Data-Reso}, the unresolved PACS and SPIRE transitions used are listed in Tables \ref{table:Data-Unreso-PACS} and \ref{table:Data-Unreso-SPIRE}.
We discuss these data below.

%-----------------------------------------------------------------------------------------------------------
\subsection{APEX}				\label{subsect:Data-APEX}
%-----------------------------------------------------------------------------------------------------------

We obtained a spectral scan of R Dor with the Atacama Pathfinder EXperiment (APEX; \citeauthor{Gusten2006}~\citeyear{Gusten2006}) using the FLASH$^+$ \citep{Klein2014} and APEX-1 \citep{Vassilev2008} instruments during the period November 11-21 2014 (project ID M-094.F-0030). 
The spectral scan covered frequencies between 213 and 476 GHz, with a backend spectral resolution of 0.61 MHz.
The reduction and analysis was done with the GILDAS CLASS software package \citep{Bardeau2006}, using the same procedure as \citet{DeNutte2016}. 
First, faulty scans and spikes were removed from the data set. A first-order polynomial baseline was subtracted from each scan, which was then averaged. 
Finally, the data were rebinned to obtain a suitable signal-to-noise ratio (SNR).
The main-beam efficiencies, used to convert the antenna temperatures to main-beam temperatures, were taken from \citet{Gusten2006} and \citet{Klein2014}.

%-----------------------------------------------------------------------------------------------------------
\subsection{Herschel/HIFI}				\label{subsect:Data-HIFI}
%-----------------------------------------------------------------------------------------------------------

Within the framework of the HIFISTARS programme \citep{Menten2010}, observations have been done with the {Heterodyne Instrument for the Far Infrared} (HIFI; \citeauthor{DeGraauw2010}~\citeyear{DeGraauw2010}) in March 2010. \cite{Justtanont2012} have presented the results, identifying the $J = 16-15$ transition for SiO and the $J = 13-12$ transition for HCN. 
Further observations of R Dor were obtained during the period April-May 2012 (P.I. Leen Decin).

The reduction and line identification were performed by \cite{KhouriPhD} using the HIFI pipeline version available in HIPE 11, the \emph{Herschel} interactive pipeline. 
The script \textsc{fitBaseline}, available in the reduction tool, was used to subtract the continuum from the observed spectra.
The spectra in the two polarization directions were combined.
The data have been corrected for the updated main-beam efficiencies \citep{HIFI}.

The integrated main-beam temperatures, listed in Table \ref{table:Data-Reso}, for APEX and HIFI are obtained by integrating the observed main-beam temperature profile over velocity. 
The uncertainty on the integrated line strengths is assumed to be 20\% \citep{Maercker2016}.
For noisy lines, we have used a line-profile fit to determine their integrated line strength.
The initial fit used a soft-parabola profile (see, e.g., Eq. 1 in \citeauthor{DeBeck2010}~\citeyear{DeBeck2010}). If the line shape deviates significantly from a soft-parabola profile, or if the line is very noisy, we have used a Gaussian line profile instead.
All spectrally resolved rotational transitions of SiO with $J_{\mathrm{up}} > 16$ are noisy (SNR less than 15), where $J_{\mathrm{up}}$ is the upper level $J$ of the transition. These are fitted by a Gaussian profile. 
The $J = 1-0$, $7-6 $, and $13-12 $ transitions of HCN are noisy as well. The $J = 1-0$ and $13-12$ transitions are fitted by a soft-parabola profile, the $J = 7-6$ transition is fitted by a Gaussian profile.

\begin{table}
    \caption{Spectrally resolved rotational transitions in the ground vibrational state of CO, SiO, and HCN used in this paper. The rest frequency, upper energy level, telescope, integrated main-beam temperature $\int{T_{\text{mb}}\ \mathrm{d}v}$, the noise rms of the continuum $\sigma$, and main-beam efficiency $\eta_{\mathrm{mb}}$ are listed as well. 
    }
    \begin{threeparttable}
    \centering
	\resizebox{\columnwidth}{!}{%
    \begin{tabular}{l r r c c c c c}
    \hline \hline 
    \noalign{\smallskip}
    Rotational	& Frequency	& $E_{\text{upper}}$ & Telescope & $\int{T_{\text{mb}}\ \mathrm{d}v} $ & $\sigma$ & $\eta_{\mathrm{mb}}$ & Ref.	\\
    transition  &	[GHz]	& 		[cm$^{-1}$]				 &	         &         [K km s$^{-1}$]           & [K]	&		   &        \\
    \noalign{\smallskip}
    \hline
	\noalign{\smallskip}
    \multicolumn{8}{c}{\textbf{CO}} \\
    \noalign{\smallskip}
    $J = 1-0$      & 115.270       & 3.85          & SEST  & 4.83    & 0.03     & 0.70  & 5    \\
    $J = 2-1$      & 230.537       & 11.54         & SEST  & 44.14   & 0.14     & 0.50  & 5    \\
    $J = 2-1$      & 230.537       & 11.54         & APEX  & 36.94   & 0.04     & 0.75  & 1    \\
    $J = 3-2$      & 345.796       & 23.07         & SEST  & 62.41   & 0.29     & 0.25  & 5    \\
    $J = 3-2$      & 345.796       & 23.07         & APEX  & 59.25   & 0.03     & 0.73  & 1    \\
    $J = 4-3$      & 461.042       & 38.45         & APEX  & 72.31   & 0.09     & 0.60  & 1    \\
    $J = 6-5$      & 691.474       & 80.74         & HIFI  & 17.50   & 0.02     & 0.65  & 1    \\
    $J = 7-6$      & 806.652       & 107.64        & HIFI  & 16.60   & 0.01     & 0.625 & 1    \\
    $J = 10-9$     & 1151.986      & 211.40        & HIFI  & 18.50   & 0.06     & 0.59  & 4    \\
    $J = 13-12$    & 1496.924      & 349.70        & HIFI  & 14.62   & 0.12     & 0.58  & 3    \\
    $J = 16-15$    & 1841.347      & 522.48        & HIFI  & 22.76   & 0.13     & 0.58  & 4    \\
    \noalign{\smallskip}
    \multicolumn{8}{c}{\textbf{SiO}} \\
    \noalign{\smallskip}
	$J = 2-1 $     & 86.847        & 4.35    & SEST  	& 9.56     & 0.02   &  0.75    & 2    \\
	$J = 3-2 $     & 130.269       & 8.69    & SEST  	& 23.96    & 0.05   &  0.65    & 2    \\
	$J = 5-4 $     & 217.105       & 21.73   & SEST  	& 29.10    & 0.14   &  0.55    & 2    \\
	$J = 5-4 $     & 217.105       & 21.73   & APEX  	& 35.72    & 0.02   &  0.75    & 1    \\
	$J = 6-5 $     & 260.518       & 30.42   & SEST  	& 34.57    & 0.20   &  0.45    & 2    \\
	$J = 6-5 $     & 260.518       & 30.42   & APEX  	& 42.88    & 0.03   &  0.75    & 1    \\
	$J = 7-6 $     & 303.927       & 40.55   & APEX  	& 41.32    & 0.01   &  0.73    & 1    \\
	$J = 8-7 $     & 347.331       & 52.14   & APEX  	& 41.32    & 0.05   &  0.73    & 1    \\
	$J = 9-8 $     & 390.729       & 65.17   & APEX  	& 52.20    & 0.04   &  0.60    & 1    \\
	$J = 10-9 $    & 434.120       & 79.65   & APEX 	& 58.28    & 0.24   &  0.60    & 1    \\
	$J = 12-11 $   & 520.881       & 112.96  & HIFI  	& 7.18     & 0.03   &  0.62    & 1    \\
	$J = 14-13 $   & 607.608       & 152.05  & HIFI  	& 6.99     & 0.07   &  0.62    & 1    \\
	$J = 16-15 $   & 694.294       & 196.92  & HIFI  	& 6.35     & 0.05   &  0.65    & 3    \\
	$J = 18-17 $   & 780.934       & 247.57  & HIFI  	& 7.01     & 0.12   &  0.65    & 1    \\
	$J = 25-24 $   & 1083.731      & 470.33  & HIFI 	& 6.69     & 0.12   &  0.635   & 1    \\
	$J = 28-27 $   & 1213.247      & 587.43  & HIFI  	& 5.92     & 0.10   &  0.59    & 1    \\
	$J = 35-34 $   & 1514.714      & 910.97  & HIFI  	& 6.58     & 0.14   &  0.58    & 1    \\
	$J = 38-37 $   & 1643.551      & 1071.14 & HIFI  & 4.58     & 0.16   &  0.58    & 1    \\
    \noalign{\smallskip}
    \multicolumn{8}{c}{\textbf{HCN}} \\
    \noalign{\smallskip}
    $J = 1-0 $     & 88.632        & 4.25    & SEST     & 0.42 & 0.01   &  0.75    & 4    \\
	$J = 3-2 $     & 265.886       & 25.52   & APEX     & 6.96 & 0.03   &  0.75    & 1    \\
%	$J = 4-3 $     & 354.505       & 42.54   & APEX     & 6.67 & 0.03   &  0.73    & 1    \\
	$J = 4-3 $     & 354.505       & 42.54   & APEX     & 6.03 & 0.03   &  0.73    & 1    \\
	$J = 7-6 $     & 620.304       & 119.09  & HIFI     & 2.24 & 0.07   &  0.62    & 1    \\
	$J = 13-12 $   & 1151.449      & 386.94  & HIFI     & 2.21 & 0.07   &  0.59    & 3    \\
    \hline
    \end{tabular}%
    }
    \begin{tablenotes}
    \footnotesize
    \item References. (1) New data (see Sect. \ref{subsect:Data-APEX} and \ref{subsect:Data-HIFI}); \\
    (2) \citet{GonzalezDelgado2003}; (3) \citet{Menten2010}; \\
    (4) \citet{Olofsson1998}; (5) \citet{Kerschbaum1999} 
    \end{tablenotes}
    \end{threeparttable}
    \label{table:Data-Reso}
\end{table}

\begin{table}
\centering
    \caption{Spectrally unresolved PACS rotational transitions in the ground vibrational state of CO, SiO, and HCN used in this paper. The rest wavelength and integrated line strength $F_\mathrm{int}$ are listed. The percentages in parentheses indicate the uncertainty on $F_\mathrm{int}$, which includes both the fitting uncertainty and the PACS absolute flux uncertainty of 15\%. 
  All blends are excluded from the modelling procedure.}
    \begin{threeparttable}
    \centering
    {\footnotesize
    \begin{tabular}{l l l c r}
    \hline \hline 
    \noalign{\smallskip}

PACS&Molecule&Rotational&$\lambda_0$&\multicolumn{1}{c}{$F_\mathrm{int}$} \\
band&&transition&[$\mu$m]&\multicolumn{1}{c}{[W m$^{-2}$]} \\\hline
\noalign{\smallskip}
R1B&SiO&$J=37 - 36$&187.30&1.16e-16 (22.2\%)\\
&CO&$J=14 - 13$&186.00&2.43e-16 (16.8\%)\\
&SiO&$J=38 - 37$&182.38&1.99e-16 (18.7\%)\\
&SiO&$J=39 - 38$&177.75&\tablefootmark{$\star$}1.61e-16 (19.1\%)\\
&CO&$J=15 - 14$&173.63&2.94e-16 (21.1\%)\\
&SiO&$J=40 - 39$&173.34&7.67e-17 (54.5\%)\\
&CO&$J=16 - 15$&162.81&3.88e-16 (16.2\%)\\
&HCN&$J=21 - 20$&161.35&\tablefootmark{$\star$}2.02e-16 (19.0\%)\\
&CO&$J=17 - 16$&153.27&3.77e-16 (19.3\%)\\
&CO&$J=18 - 17$&144.78&\tablefootmark{$b$}4.61e-16 (16.6\%)\\\hline
\noalign{\smallskip}
R1A&CO&$J=18 - 17$&144.78&\tablefootmark{$c$}5.06e-16 (18.7\%)\\
&CO&$J=19 - 18$&137.20&\tablefootmark{$\star$}4.86e-16 (19.6\%)\\
&HCN&$J=25 - 24$&135.63&\tablefootmark{$a$}1.81e-16 (30.3\%)\\
&HCN&$J=26 - 25$&130.44&\tablefootmark{$\star$}4.66e-16 (18.8\%)\\
&CO&$J=20 - 19$&130.37&Blended\\
&CO&$J=21 - 20$&124.19&3.77e-16 (19.3\%)\\
&CO&$J=22 - 21$&118.58&\tablefootmark{$b$}1.51e-15 (15.7\%)\\
&CO&$J=23 - 22$&113.46&\tablefootmark{$a$}3.25e-15 (15.6\%)\\
&CO&$J=24 - 23$&108.76&\tablefootmark{$a$}5.80e-16 (24.9\%)\\\hline
\noalign{\smallskip}
B2A&CO&$J=36 - 35$&72.84&\tablefootmark{$b$}3.41e-15 (16.3\%)\\
&CO&$J=38 - 37$&69.07&2.81e-16 (31.7\%)\\\hline
\noalign{\smallskip}
B2B&CO&$J=28 - 27$&93.35&\tablefootmark{$\star$}1.12e-15 (15.9\%)\\
&CO&$J=29 - 28$&90.16&2.29e-16 (74.6\%)\\
&CO&$J=31 - 30$&84.41&\tablefootmark{$b$}5.76e-16 (20.3\%)\\
&CO&$J=36 - 35$&72.84&3.46e-15 (17.2\%)\\    \hline
    \end{tabular}%
    }
    \begin{tablenotes}[flushleft]
    \footnotesize
    \item $^{(\star)}$ Lines flagged for potential line blends (see Sect. \ref{subsect:Data-PACS}). 
    Transitions that might contribute to the blended line are listed 
    immediately below the flagged transition. $^{(a)}$ Identified blends of strong 
    water transitions. $^{(b)}$ Identified blend with vibrationally exctited SiO. 
    $^{(c)}$ Identified blend with H$_2$S.
    \end{tablenotes}
    \end{threeparttable}
    \label{table:Data-Unreso-PACS}
\end{table}

\begin{table}
\centering
    \caption{Spectrally unresolved SPIRE rotational transitions in the ground vibrational state of CO, SiO, and HCN used in this paper. The rest wavelength and integrated line strength $F_\mathrm{int}$ are listed. The percentages in parentheses indicate the uncertainty on $F_\mathrm{int}$, which includes both the fitting uncertainty and the SPIRE absolute flux uncertainty of 15\%. 
%FWHM x 1.3, extra's
    }
    \begin{threeparttable}
    \centering
	{\footnotesize
    \begin{tabular}{l l l c r}
    \hline \hline 
    \noalign{\smallskip}

SPIRE&Molecule&Rotational&$\lambda_0$&\multicolumn{1}{c}{$F_\mathrm{int}$} \\
band&&transition&[$\mu$m]&\multicolumn{1}{c}{[W m$^{-2}$])} \\\hline
\noalign{\smallskip}
SLW&CO&$J=4 - 3$&650.25&7.61e-17 (16.5\%)\\
&SiO&$J=11 - 10$&627.75&3.10e-17 (22.5\%)\\
&SiO&$J=12 - 11$&575.71&3.81e-17 (20.1\%)\\
&SiO&$J=13 - 12$&531.35&5.10e-17 (18.1\%)\\
&CO&$J=5 - 4$&520.23&9.90e-17 (15.9\%)\\
&SiO&$J=14 - 13$&493.34&4.75e-17 (18.4\%)\\
&HCN&$J=7 - 6$&483.30&\tablefootmark{$\star$}1.81e-16 (15.3\%)\\
&SiO&$J=15 - 14$&460.62&3.51e-17 (19.2\%)\\
&CO&$J=6 - 5$&433.56&1.35e-16 (15.5\%)\\
&SiO&$J=16 - 15$&431.78&3.92e-17 (18.6\%)\\
&SiO&$J=17 - 16$&406.50&5.07e-17 (18.0\%)\\
&SiO&$J=18 - 17$&383.88&6.22e-17 (17.1\%)\\
&CO&$J=7 - 6$&371.65&\tablefootmark{$\star$}1.71e-16 (15.4\%)\\
&SiO&$J=19 - 18$&363.77&6.49e-17 (17.4\%)\\
&SiO&$J=20 - 19$&345.54&7.64e-17 (18.2\%)\\
&SiO&$J=21 - 20$&329.16&7.76e-17 (16.4\%)\\
&CO&$J=8 - 7$&325.23&\tablefootmark{$\star$}1.78e-16 (15.3\%)\\
&SiO&$J=22 - 21$&314.27&8.59e-17 (16.2\%)\\
&SiO&$J=23 - 22$&300.57&7.73e-17 (18.2\%)\\
\hline
\noalign{\smallskip}
SSW&SiO&$J=22 - 21$&314.27&8.03e-17 (20.0\%)\\
&SiO&$J=23 - 22$&300.57&8.66e-17 (19.1\%)\\
&CO&$J=9 - 8$&289.12&2.02e-16 (15.8\%)\\
&SiO&$J=24 - 23$&288.10&\tablefootmark{$\star$}9.11e-17 (19.4\%)\\
&HCN&$J=12 - 11$&282.03&\tablefootmark{$\star$}3.77e-17 (87.7\%)\\
&SiO&$J=25 - 24$&276.63&6.24e-17 (20.8\%)\\
&SiO&$J=26 - 25$&266.03&8.10e-17 (19.1\%)\\
&HCN&$J=13 - 12$&260.36&\tablefootmark{$\star$}3.22e-16 (16.0\%)\\
&CO&$J=10 - 9$&260.24&Blended\\
&SiO&$J=27 - 26$&256.21&\tablefootmark{$\star$}8.30e-17 (23.4\%)\\
&SiO&$J=28 - 27$&247.10&1.18e-16 (18.1\%)\\
&SiO&$J=29 - 28$&238.61&7.74e-17 (19.8\%)\\
&CO&$J=11 - 10$&236.61&2.30e-16 (15.6\%)\\
&SiO&$J=31 - 30$&223.31&1.07e-16 (17.8\%)\\
&CO&$J=12 - 11$&216.93&2.41e-16 (42.3\%)\\
&SiO&$J=32 - 31$&216.36&1.17e-16 (18.2\%)\\
&SiO&$J=33 - 32$&209.82&1.15e-16 (23.2\%)\\
&SiO&$J=34 - 33$&203.71&9.65e-17 (18.3\%)\\
&CO&$J=13 - 12$&200.27&2.38e-16 (15.6\%)\\
&SiO&$J=35 - 34$&197.90&1.05e-16 (17.8\%)\\\hline
    \end{tabular}%
    }
    \begin{tablenotes}[flushleft]
    \footnotesize
    \item $^{(\star)}$ Lines flagged for potential line blends (see Sect. \ref{subsect:Data-PACS}). Transitions that might contribute to the blended line are listed immediately below the flagged transition. 
    $^{(a)}$ Identified blends of strong water transitions. 
    \end{tablenotes}
    \end{threeparttable}
    \label{table:Data-Unreso-SPIRE}
\end{table}

%-----------------------------------------------------------------------------------------------------------
\subsection{Herschel/PACS}			\label{subsect:Data-PACS}
%-----------------------------------------------------------------------------------------------------------

Within the framework of the MESS programme (\citeauthor{Groenewegen2011}~\citeyear{Groenewegen2011}), observations have been done with the {Photodetector Array Camera and Spectrometer} (PACS; \citeauthor{Poglitsch2010}~\citeyear{Poglitsch2010}). The spectra were taken on June 5 2010 and January 12 2011 (Obs. IDs 1342197795 and 1342197794).
R Dor was also observed on 24 September 2011 (Obs. IDs 1342229706 and 1342229708) for calibration purposes. 
We have opted to use the spectrum obtained on June 5 2010 (Obs. ID 1342197795), following \citet{KhouriPhD}.
The calibration was performed by telescope background normalization with HIPE 13 using calibration set 69 and applying a point-source correction. 

Table \ref{table:Data-Unreso-PACS} lists the PACS molecular line transitions used in this paper. The line strengths were measured by fitting a Gaussian on top of a continuum. The uncertainties include the fitting uncertainty of the Gaussian fits and the adopted absolute-flux-calibration uncertainty of 15\%  \citep{Lombaert2016}.
Measured transitions are flagged as blends if multiple transitions were identified with a central wavelength within the full width at half maximum (FWHM) of the fitted line and/or if the FWHM of the fitted Gaussian is at least 20\% larger than the FWHM of the PACS spectral resolution \citep{Lombaert2016}.

%-----------------------------------------------------------------------------------------------------------
\subsection{Herschel/SPIRE}			\label{subsect:Data-SPIRE}
%-----------------------------------------------------------------------------------------------------------

The MESS programme \citep{Groenewegen2011} also included observations with the {Spectral and Photometric Imaging Receiver Fourier-Transform Spectrometer} (SPIRE FTS; \citeauthor{Griffin2010}~\citeyear{Griffin2010}). The spectra were taken on October 27 2012 (Obs. ID 1342245114). 
The reduction and line identification was performed by \cite{KhouriPhD} using the HIPE SPIRE FTS pipeline version 11. R Dor was assumed to be a point source within the SPIRE beam. 
The integrated line strengths were measured using the script \textsc{Spectrometer Line Fitting} available in HIPE. This script simultaneously fits a power-law to the continuum and a cardinal sine function to the molecular lines in the unapodized spectra.

Table \ref{table:Data-Unreso-SPIRE} lists the SPIRE molecular line transitions used in this paper. The uncertainties listed include the fitting uncertainties of the cardinal sine fits and our adopted absolute-flux-calibration uncertainty of 15\%.
Blends are flagged using the same method as for the PACS spectra. 
However, considering a line as blended when the FWHM of the fitted sinc function is at least 20\% larger than the FWHM of the SPIRE spectral resolution excluded those that agreed with measurements of the same molecular line obtained with a different telescope. We have therefore opted to increase the factor to 30\%, which allows these lines to be included.

%-----------------------------------------------------------------------------------------------------------
\subsection{Literature molecular data}			\label{subsect:Data-lit}
%-----------------------------------------------------------------------------------------------------------

Additional molecular transitions have been taken from the literature.
\cite{GonzalezDelgado2003} have measured the $J = 2-1$, $3-2$ , $5-4$, and $6-5$ molecular transitions of SiO with the Swedish-ESO 15 m Submillimetre Telescope (SEST). The $J = 2-1$, $5-4$, and $6-5$ molecular transitions were observed in December 1992, the  $J = 3-2$ molecular transition in August 2001. 
\cite{Olofsson1998} have measured the  $J = 2-1$ molecular transition of HCN with SEST 
over the years 1987 to 1997.

%-----------------------------------------------------------------------------------------------------------
\subsection{Infrared spectrum}                    \label{subsect:Data-spectra}
%-----------------------------------------------------------------------------------------------------------

R Dor was observed with the short-wavelength spectrometer (SWS, \citeauthor{DeGraauw1996}~\citeyear{DeGraauw1996}) onboard the Infrared Space Observatory (ISO, \citeauthor{Kessler1996}~\citeyear{Kessler1996}), which covers the spectral range from 2.4 $\mu$m to 45.4 $\mu$m. 
The reduced spectrum used in this paper was retrieved from the \citet{Sloan2003} database. 
The observation (identification number 58900918) was done in observing mode SWS01 with scan speed 1 on June 27 1996, providing a low-resolution (R $\sim$ 20000) full grating scan.
%REMARK The observation, with identification number 58900918, took place on June 27 1996.

%-----------------------------------------------------------------------------------------------------------
\subsection{Photometric data}                    \label{subsect:Data-photo}
%-----------------------------------------------------------------------------------------------------------

Photometric data was retrieved using the VizieR catalogue access tool. We have excluded data from the DENIS catalogue as the source was saturated. 
Data from the WISE catalogue was excluded as well because of indications that the source might be spatially resolved.
All photometric data used are listed in Appendix \ref{app:PhotometricData}.

%%%%%%%%%%%%%%%%%%%%%%%%%%%%%%%%%%%%%%%%%%%%%%%%%%%%%%%
\section{Methodology}							\label{sect:Methodology}
%%%%%%%%%%%%%%%%%%%%%%%%%%%%%%%%%%%%%%%%%%%%%%%%%%%%%%%

As the CSE is composed of both dust and gas-phase species, information on both components should be coupled in order to obtain a full understanding of the entire CSE. 
In Sect. \ref{subsect:Meth-MCMax} and \ref{subsect:Meth-GASTRoNOoM}, we elaborate on the continuum and line radiative-transfer codes used. The approach used for solving the continuum and line radiative transfer in a coherent manner is explained in Sect. \ref{subsect:Meth-Approach}.
Finally, we expand on the criteria used to determine the goodness-of-fit of our models in Sect. \ref{subsect:Meth-GoodnessOfFit}.

%----------------------------------------------------------------------------------------------------------
\subsection{Continuum radiative transfer}		\label{subsect:Meth-MCMax}
%----------------------------------------------------------------------------------------------------------

The continuum radiative transfer was performed using the Monte Carlo radiative-transfer code MCMax \citep{Min2009}. The code predicts the dust temperature stratification and the emergent thermal infrared continuum. A spherically symmetric dust envelope was assumed throughout the modelling. 
The distance to R Dor is taken to be 59 pc \citep{Knapp2003,KhouriPhD}. 

The dust component of the outflow consists of amorphous Al$_2$O$_3$, Ca$_2$Mg$_{0.5}$Al$_2$Si$_{1.5}$O$_7$ (melilite), and metallic iron. The dust opacity is calculated for particle shapes represented by a distribution of hollow spheres with a filling factor of 0.8 \citep{Min2003}. 
The optical constants of amorphous Al$_2$O$_3$ were measured by \citet{Koike1995}, those of melilite and metallic iron by \citet{Mutschke1998} and \citet{Henning1996} respectively.
The optical constants of melilite are only available down to $\sim$ 6 $\mu$m. 
For shorter wavelengths, they are approximated using the optical constants of MgSiO$_3$, which like melilite does not contain any iron.
It is the iron content that has the strongest effect on the absorption at short wavelengths.
The dust grain population is assumed to follow the size distribution of \citet{Mathis1977},
\begin{equation}
n_d(a,r)\ \mathrm{d}a = A(r)\ a^{-3.5}\ \mathrm{n_H}(r)\ \mathrm{d}a, 
\end{equation}
with n$_\mathrm{H}$ the total hydrogen number density, $a$ the radius of the spherical dust grain, and $A(r)$ the abundance scale factor giving the number of dust particles with respect to hydrogen. 
For Al$_2$O$_3$, the minimum and maximum grain size in the distribution are taken to be 0.001 $\mu$m and 0.3 $\mu$m. For metallic iron and melilite, the minimum and maximum grain sizes are 0.29 $\mu$m and 0.31 $\mu$m.

%----------------------------------------------------------------------------------------------------------
\subsection{Line radiative transfer}				\label{subsect:Meth-GASTRoNOoM}
%----------------------------------------------------------------------------------------------------------

Spectral line profiles were computed using the non-local thermal equilibrium (non-LTE) code GASTRoNOoM \citep{Decin2006, Decin2010}. In this model, the outflow is assumed to be spherically symmetric with a smooth density distribution, implying that neither small-scale (e.g. clumps) nor large-scale (e.g. spirals, disks) density features can be accounted for.
When calculating the line profiles, the beam shape of the used telescope is taken into account such that observed and modelled line profiles can be compared directly.
The molecular data are described in \citet{Decin2010}.

The velocity profile of the outflow is parametrised using a $\beta$-type law,
\begin{equation}
\upsilon(r) = \upsilon_o + \left(\upsilon_\infty - \upsilon_o\right) \left(1 - \frac{r_o}{r}\right)^\beta,
\end{equation}
with $r$ the distance to the star. In this equation, $v_o$ is the velocity at radius $r_o$ at which the wind is launched and is set to the local sound speed. The velocity profile for $r < r_o$, i.e. the subsonic region, is fixed and characterised by a $\beta$-type law with $\beta = 0.5$ \citep{Decin2006}. %, a typical value for an optically thin wind
The temperature structure of the gas is approximated using a power law,
\begin{equation}
T(r) = T_* \left( \frac{R_*}{r} \right)^{-\epsilon}.
\end{equation}

The CO photodissociation radius is prescribed by the formalism of \citet{Mamon1988}.
The abundance profiles of SiO and HCN assume a constant initial abundance w.r.t. H$_2$ up to a certain distance from the star, $R_{\mathrm{decl}}$. From that distance onwards, the abundance declines according to a Gaussian profile to mimic the gradual photodissociation of the molecule.
The Gaussian profile is given by 
\begin{align}
f(r) &= f_0\ \mathrm{exp}\left( -\left( \frac{r}{r_\mathrm{e}} \right)^2 \right)  && \text{for $R > R_{\mathrm{decl}}$,}
\end{align}
with $f_0$ the initial abundance and $r_\mathrm{e}$ the e-folding radius. 
The profile is calculated as centered on the star, and then shifted to $R_{\mathrm{decl}}$. In order to avoid any discontinuities, the  profile is scaled to the abundance at $R_{\mathrm{decl}}$.
Note that the initial abundance is not equal to the abundance at the stellar surface, but rather the abundance probed by the highest rotational transitions of the molecule. 
The abundance profiles of SiO and HCN are hence characterised by three parameters: \emph{(i)} the initial abundance $f_0$ with respect to H$_2$; \emph{(ii)} the maximum radius of the constant abundance profile $R_{\mathrm{decl}}$ in $R_*$; and \emph{(iii)} the e-folding radius of the Gaussian profile $r_\mathrm{e}$ in cm.

%----------------------------------------------------------------------------------------------------------
\subsection{Modelling approach}						\label{subsect:Meth-Approach}
%----------------------------------------------------------------------------------------------------------

The continuum and line radiative transfer is solved using a five-step approach\footnote{\url{https://github.com/robinlombaert/ComboCode}} based on \citet{Lombaert2013}, where we iterate between MCMax and GASTRoNOoM:
\begin{enumerate}
\item Using MCMax, we obtain an initial estimate of the dust composition, dust temperature and dust mass-loss rate by modelling the dust IR continuum.
\item The dust extinction efficiencies, grain temperatures, and dust mass-loss rate obtained from MCMax are incorporated in the GASTRoNOoM iteration.
Using GASTRoNOoM, the kinematics and thermodynamics of the gas shell are calculated. 
This step provides us with a model for the momentum transfer from dust to gas, i.e. a dust velocity profile and the drift velocity between dust and gas.
\item The dust velocity profile together with the given dust mass-loss rate are used to compute a new dust density profile. The IR continuum model is updated with MCMax using the new dust density profile.
\item The thermodynamics of the wind are updated with GASTRoNOoM using the updated dust parameters.
\item Line radiative transfer is performed by GASTRoNOoM and the line profiles are calculated.
\end{enumerate}

%----------------------------------------------------------------------------------------------------------
\subsection{Criteria used for the goodness-of-fit to molecular data}	\label{subsect:Meth-GoodnessOfFit}
%----------------------------------------------------------------------------------------------------------

Two types of uncertainties determine the error on the observed molecular transitions. The systematic error with variance $\sigma^2_{\mathrm{abs}}$ arises from e.g. uncertainties in the correction for the erratic fluctuations in atmosphere emission and calibration uncertainties.
The random error with variance $\sigma^2_{\mathrm{err}}$ accounts for the statistical variation of the measured line flux within a certain bin \citep{Decin2007}. 
When determining the goodness-of-fit of a model to the molecular data, both uncertainties need to be taken into account. 
For all resolved and most unresolved observed molecular lines used in this paper, the random error is smaller than the systematic error. 

In order to determine whether a model yields a good fit to the molecular data, we use two criteria: the integrated line strength and the log-likelihood function. 
The integrated line strength is applicable to both resolved and unresolved molecular lines. The criterion for a good fit of the model to the data is that the integrated line strength of all lines lies within the uncertainty of the integrated line strength of the data. This uncertainty is mainly determined by the systematic error on the observation \citep{Skinner1999}.
The log-likelihood function \citep{Decin2007} indicates whether the shape of the modelled line agrees with that of the observed molecular line. It is hence not applicable to noisy resolved molecular lines and unresolved molecular lines. 
Since the statistical error is much smaller than the systematic error for all observed resolved molecular lines, the line profile shape is of greater statistical significance than the integrated line strength. 
It is also the shape of resolved line profiles that contains diagnostics of their line formation regions \citep{Decin2006}.
When calculating the log-likelihood function, the observed molecular line is first normalised to its integrated line strength. The modelled line profile is then scaled to the normalised observed molecular line to ensure that all models considered are treated equally. 
Statistically, the scaling is just an additional parameter, accounting for calibration uncertainties. The log-likelihood function is used to determine the 95\% confidence intervals of the parameters describing the abundance profile.

Ideally, a model satisfies both criteria: the integrated line strength of the modelled line lies within the uncertainty of the integrated line strength of the observed molecular line, and the modelled line profile satisfies the log-likelihood function.
These criteria also allow us to find model degeneracies, i.e. models with equal statistical significance.

%%%%%%%%%%%%%%%%%%%%%%%%%%%%%%%%%%%%%%%%%%%%%%%%%%%%%%%
\section{Results}					\label{sect:Results}
%%%%%%%%%%%%%%%%%%%%%%%%%%%%%%%%%%%%%%%%%%%%%%%%%%%%%%%

In Sect. \ref{subsect:Results-EnvModel} we describe the kinematical and thermodynamical envelope model of R Dor, which is largely based on the results of \citet{KhouriPhD}.
Using this model, we have determined the abundances of SiO and HCN throughout the wind, thus creating abundance profiles of these molecules valid for the regions probed by the observational data.
The SiO and HCN abundance profiles are discussed in Sect. \ref{subsect:Results-AbunProf-SiO} and \ref{subsect:Results-AbunProf-HCN}.
Molecular abundances are given relative to the molecular hydrogen abundance H$_2$.

%----------------------------------------------------------------------------------------------------------
\subsection{The envelope model}        \label{subsect:Results-EnvModel}
%----------------------------------------------------------------------------------------------------------

The adopted envelope model used is based on that found by \citet{KhouriPhD}, who used the same radiative transfer codes and largely the same methodology.
In this section, we describe the modelling of both the dusty and gaseous components, and expand on the differences between our envelope model and that of \citet{KhouriPhD}. 
The parameters of both the dust and gas-phase component of the envelope model can be found in Table \ref{table:Model-parameters}.

\begin{table}[b]
	\caption{Best-fit parameters for the envelope model. Listed are the assumed distance $D$, the black-body stellar effective temperature $T_{\mathrm{bb}}$, the stellar luminosity $L_*$,  the gas and dust mass-loss rates, $\dot{M}_{\mathrm{gas}}$ and $\dot{M}_{\mathrm{dust}}$, the maximum expansion velocity $\upsilon_\infty$, the radius of the onset of the acceleration of the wind $r_o$,
	 the inner radius of the silicate emission $R_{\mathrm{Sil}}$,
	the exponent of the temperature power law $\epsilon$, the exponent of the velocity profile $\beta$, and the drift and turbulent velocities $\upsilon_{\mathrm{drift}}$ and $\upsilon_{\mathrm{turb}}$
	}
	\centering
    \begin{tabular}{l l | l l }
    \hline \hline 
    \noalign{\smallskip}
    $D$ [pc]                         & 59                     & $R_{\mathrm{Sil}}$ [$R_*$]       & 60        \\
    $T_{\mathrm{bb}}$ [K]            & 2500                   & $r_o$ [$R_*$]     & 1.6       \\
    $L_*$ [$L_\odot$]                & 4500                   & $\epsilon$                         & 0.65      \\
    $\dot{M}_{\mathrm{gas}}$ [M$_\odot$ year$^{-1}$] & $1.25 \times 10^{-7}$     & $\beta$                         & 5         \\
    $\dot{M}_{\mathrm{dust}}$ [M$_\odot$ year$^{-1}$]& $4.05 \times 10^{-10}$ & $\upsilon_{\mathrm{drift}}$ [km s$^{-1}$] & 8.4     \\
    $\upsilon_\infty$ [km s$^{-1}$]         & 5.5                    & $\upsilon_{\mathrm{turb}}$ [km s$^{-1}$]  & 1 \\    
    $R_*$ [cm]                      & $2.5 \times 10^{13}$                    &                & \\
    \noalign{\smallskip}
    \hline
    \end{tabular}%
    \label{table:Model-parameters}    
\end{table}

%--.--.--.--.--.--.--.--.--.--.--.--.--.--.--.--.--.--.--.--.--.--.--.--.--.--.--.--.--.--.--.--.--.--.--.--.--.--.--
\subsubsection{Dust model}        \label{subsubsect:Results-EnvModel-Dust}
%--.--.--.--.--.--.--.--.--.--.--.--.--.--.--.--.--.--.--.--.--.--.--.--.--.--.--.--.--.--.--.--.--.--.--.--.--.--.--

The dust model was obtained by fitting the ISO-SWS spectrum.
The structure of the dust envelope is based on \citet{KhouriPhD}. 
Whereas \citet{KhouriPhD} has used a modified blackbody as the stellar spectrum (at wavelengths shorter than 0.9 $\mu$m a blackbody of 2300 K is assumed, at longer wavelengths one of 3000 K), we have used a blackbody of 2500 K as the assumed stellar spectrum. This leads to a difference in both stellar luminosity and temperature between our models.
We have also accounted for a drift velocity between gas and dust in the outflow.
As the mass-loss rate is low, it is highly unlikely that the drift veloctity is zero \citep{Kwok1975}.
The inclusion of drift gives rise to a dust mass-loss rate of 
%4.05 $\times 10^{-10}$ M$_\odot$ yr$^{-1}$ compared to the value of 1.6 $\times 10^{-10}$ M$_\odot$ yr$^{-1}$ found by \citet{KhouriPhD}.
4.1 $\times 10^{-10}$ M$_\odot$ yr$^{-1}$ compared to the value of 1.6 $\times 10^{-10}$ M$_\odot$ yr$^{-1}$ found by \citet{KhouriPhD}.
See Sect. \ref{subsect:Results-EnvModel}, Table \ref{table:Model-parameters} for all model parameters.

The structure of the dust envelope consists of a gravitationally bound dust shell (GBDS) located close to the star 
%together with a component in the stellar wind, similar to the structure of the dust envelope of W Hya \citep{Khouri2015}. 
and dust flowing out in the stellar wind, similar to the structure of the dust envelope of W Hya \citep{Khouri2015}. 
The stellar wind is launched at the outer edge of the GBDS. The existence of the GBDS was first postulated by \citet{Khouri2014a} for W Hya, and fits well within recent theoretical models of \citet{Hofner2016}.
The GBDS is considered to be static. 
The composition of the dust envelope is taken from \cite{KhouriPhD}. 
The GBDS consists of amorphous Al$_2$O$_3$ grains located between 1.7 and 1.9 $R_*$ and dominates the emission at 11-12 $\mu$m. 
The dust in the stellar wind is composed of 55\% melilite and 45\% metallic iron. 
The inner radius of the dust envelope is at $\sim$ 60 R$_*$. It is not well constrained, as there is a degeneracy between the derived dust mass-loss rate, the inner radius, and the metallic iron content \citep{KhouriPhD}.

\begin{figure}
\centering
\includegraphics[width=\columnwidth]{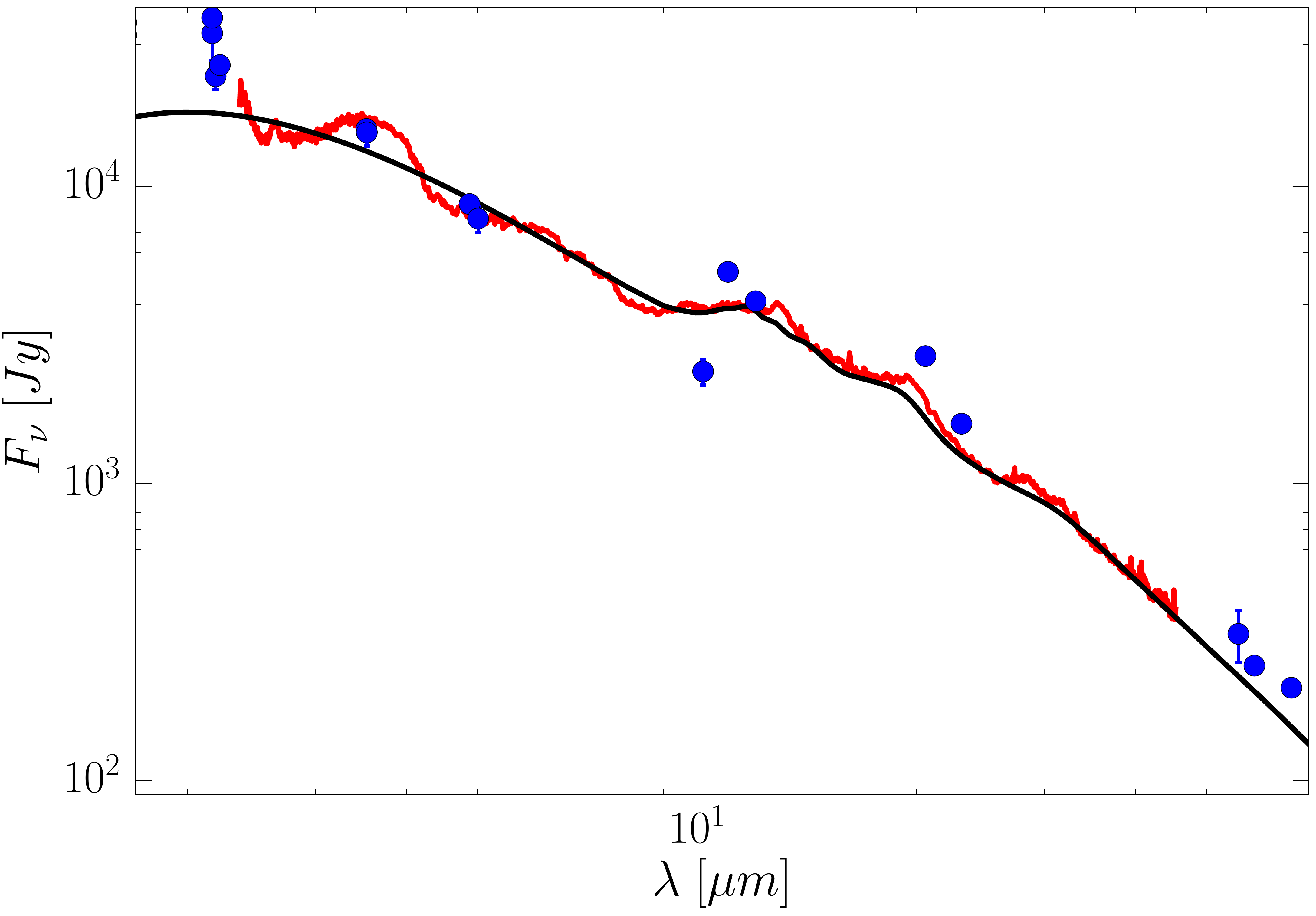}
\caption{Red: spectral energy distribution of R Dor given by the ISO-SWS spectrum. Black: the best-fitting model for the dust continuum is shown. See Sect. \ref{subsubsect:Results-EnvModel-Dust} for the dust properties of the model.}
\label{fig:Results-Dust-BestFit}
\end{figure}

Fig. \ref{fig:Results-Dust-BestFit} shows the best-fit model together with the ISO-SWS spectrum.
It reproduces the spectrum fairly well except for the 20 $\mu$m and 30 $\mu$m features, similar to the best-fit model of \cite{KhouriPhD}.
\cite{Sloan2003} suggested that these features were due to crystalline forms of silicate.
As improving the fit to the ISO-SWS spectrum lies beyond the scope of this paper, we did not attempt to identify the carrier(s) of these features.

%--.--.--.--.--.--.--.--.--.--.--.--.--.--.--.--.--.--.--.--.--.--.--.--.--.--.--.--.--.--.--.--.--.--.--.--.--.--.--
\subsubsection{Gas-phase model}        \label{subsubsect:Results-EnvModel-Gas}
%--.--.--.--.--.--.--.--.--.--.--.--.--.--.--.--.--.--.--.--.--.--.--.--.--.--.--.--.--.--.--.--.--.--.--.--.--.--.--

The gas-phase model is determined using the CO and SiO molecular data. 
Based on our SPIRE and PACS data of SiO, we derive a higher gas mass-loss rate of 
$1.3 \times 10^{-7}$ M$_\odot$ year$^{-1}$ instead of $9 \times 10^{-8}$ M$_\odot$ year$^{-1}$ as derived by \citet{KhouriPhD}. 
This value is consistent with the gas mass-loss rate found by \citet{Schoier2013} and \citet{Maercker2016}.
We have adopted a CO abundance of 3.2 $\times$ 10$^{-4}$ relative to H$_2$.
The extent of the CO envelope is fitted by the \citet{Mamon1988} abundance profile.
The $J=4-3$, $3-2$, and $2-1$ transitions are overestimated by our model and are correctly modelled by \citet{Maercker2016}. 
When using the same power-law exponent of the temperature profile as \citet{Maercker2016} ($\epsilon=0.83$), we are able to achieve a better fit to the resolved lines. However, using this value significantly underestimates all unresolved lines. Since our model achieves a better fit to more lines according to our two criteria (see Sect. \ref{subsect:Meth-GoodnessOfFit}), we chose to use a value $\epsilon=0.65$, following \citet{KhouriPhD}.
See Sect. \ref{subsect:Results-EnvModel}, Table \ref{table:Model-parameters} for all model parameters.

\begin{figure}[h]
\centering
\includegraphics[width=0.98\columnwidth]{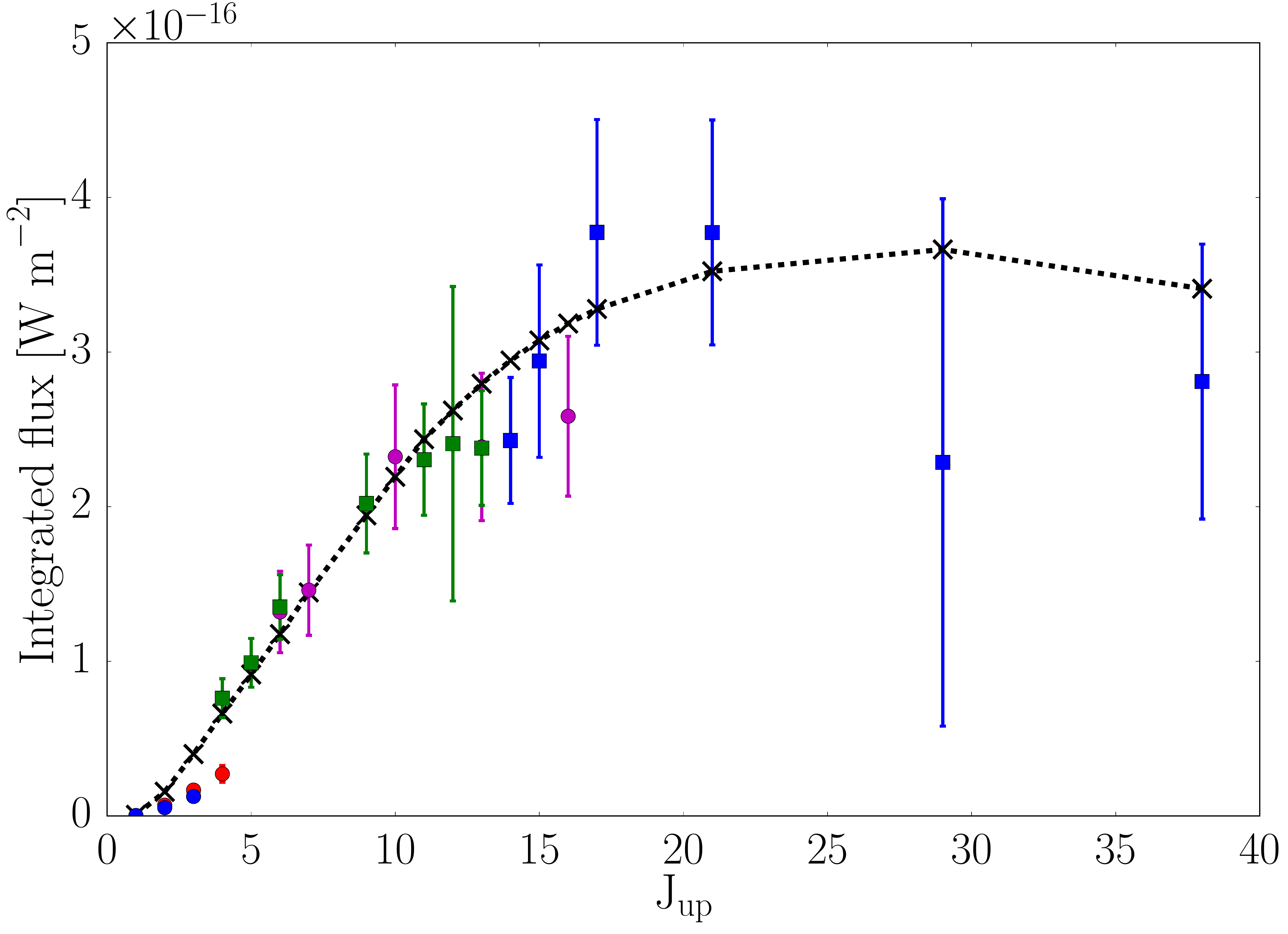}
\caption{CO model for R Dor (dashed black line). Overplotted are the integrated line strengths measured with APEX (red circles), SEST (blue circles), HIFI (purple circles), SPIRE (blue squares), and PACS (green squares). The errorbars for the resolved molecular lines correspond to an absolute error of 20\%.}
\label{fig:Results-CO-IntInt}
\end{figure}

\begin{figure*}
\centering
\includegraphics[width=0.98\textwidth]{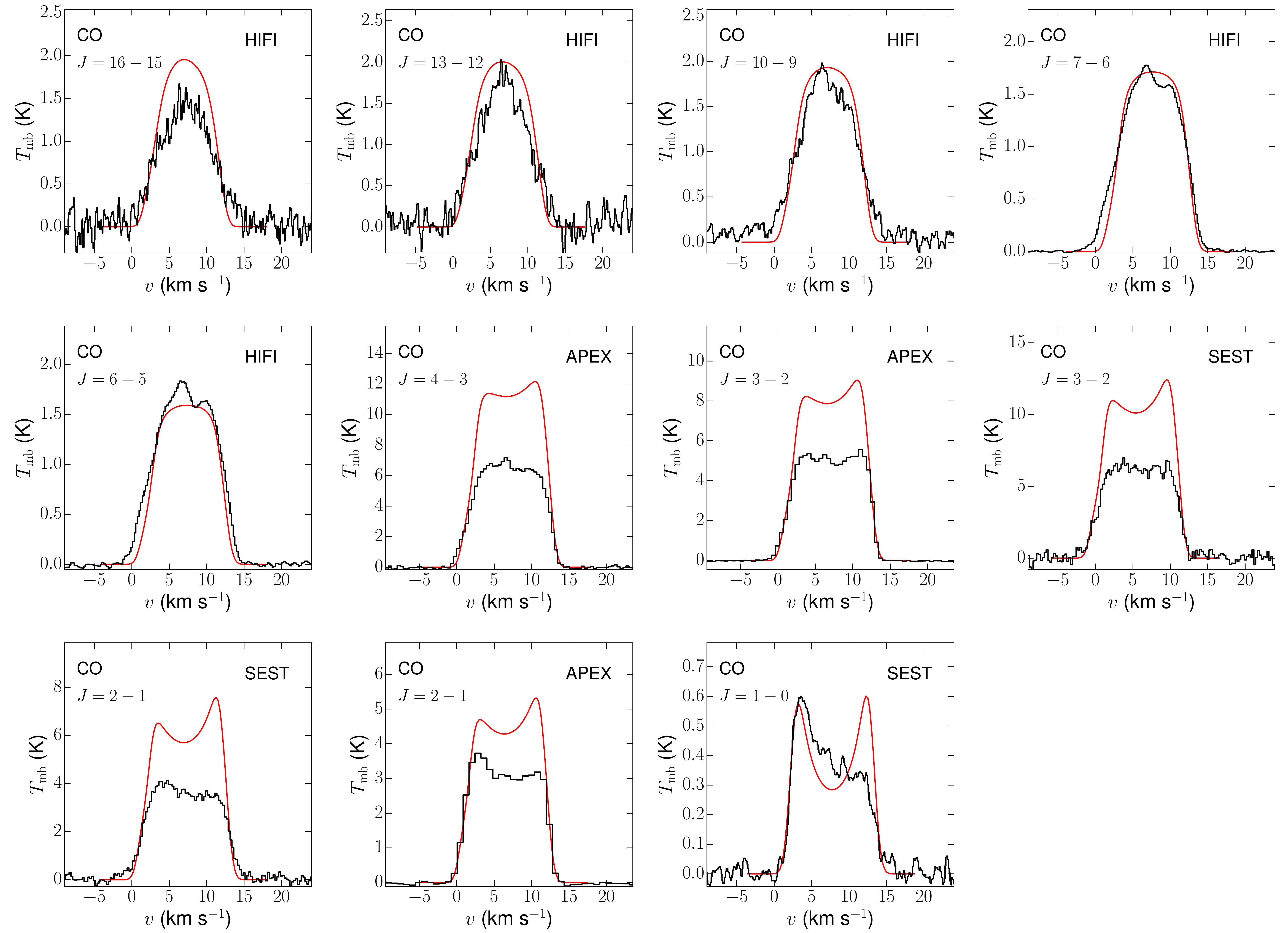}
\caption{Black: observed main-beam temperature as a function of velocity for 11 rotational transitions of CO (see Table \ref{table:Data-Reso}). The telescope and rotational quantum numbers are indicated for each line.
Red: line profiles resulting from the gas-phase model (see Sect. \ref{subsubsect:Results-EnvModel-Gas}). 
The $v_{\mathrm{LSR}}$ of R Dor is 7.5 km s$^{-1}$.
}
\label{fig:Results-CO-LPs}
\end{figure*}

Fig. \ref{fig:Results-CO-LPs} shows the 9 resolved rotational transitions in the ground vibrational state of CO that were used to determine the gas-phase model. The best-fit line profiles are shown as well. 
The integrated line strength results for the PACS and SPIRE data, together with the resolved data, are shown in Fig. \ref{fig:Results-CO-IntInt}. 
The integrated fluxes of the resolved molecular lines are converted from K km s$^{-1}$ to W m$^{-2}$ (see \citet{Schoenberg1988}, Eq. 9).
The modelled PACS spectra can be found in Appendix \ref{app:PACSSpectra}. 
All resolved rotational transitions satisfy the loglikelihood criterion. Nevertheless, only four of the eleven CO rotational transitions, namely $J = 1-0$, $6-5$, $7-6$, and $10-9$, satisfy the integrated line strength criterion. Despite this, the gas-phase model is able to fit the high-$J$ PACS and SPIRE rotational transitions of both SiO and CO simultaneously.

\begin{figure*}
\centering
\includegraphics[width=0.98\textwidth]{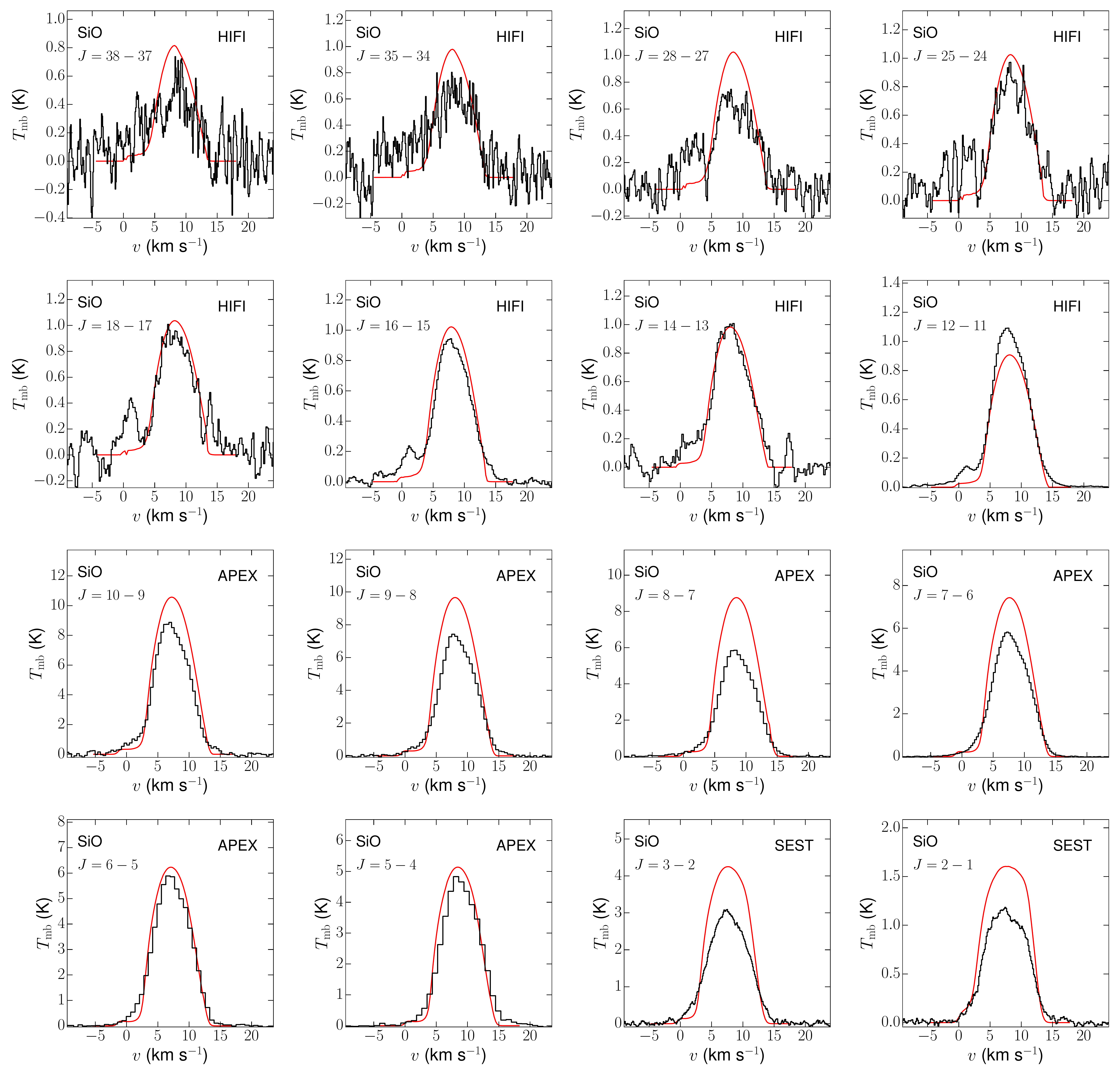}
\caption{Black: observed main-beam temperature as a function of velocity for 16 rotational transitions of SiO (see Table \ref{table:Data-Reso}). The telescope and rotational quantum numbers are indicated for each line.
Red: resulting line profiles using an abundance profile characterised by $f_0 = 6.0 \times 10^{-5}$ relative to H$_2$ and $e$-folding radius $r_e = 4 \times 10^{13}$ cm with $R_{\rm decl} = 50\ \mathrm{R}_*$.
The $\upsilon_{\mathrm{LSR}}$ of R Dor is 7.5 km s$^{-1}$.
}
\label{fig:Results-SiO-LPsGood}
\end{figure*}

%----------------------------------------------------------------------------------------------------------
\subsection{Abundance profile of SiO}			\label{subsect:Results-AbunProf-SiO}
%----------------------------------------------------------------------------------------------------------

\begin{figure}
\centering
\includegraphics[width=0.98\columnwidth]{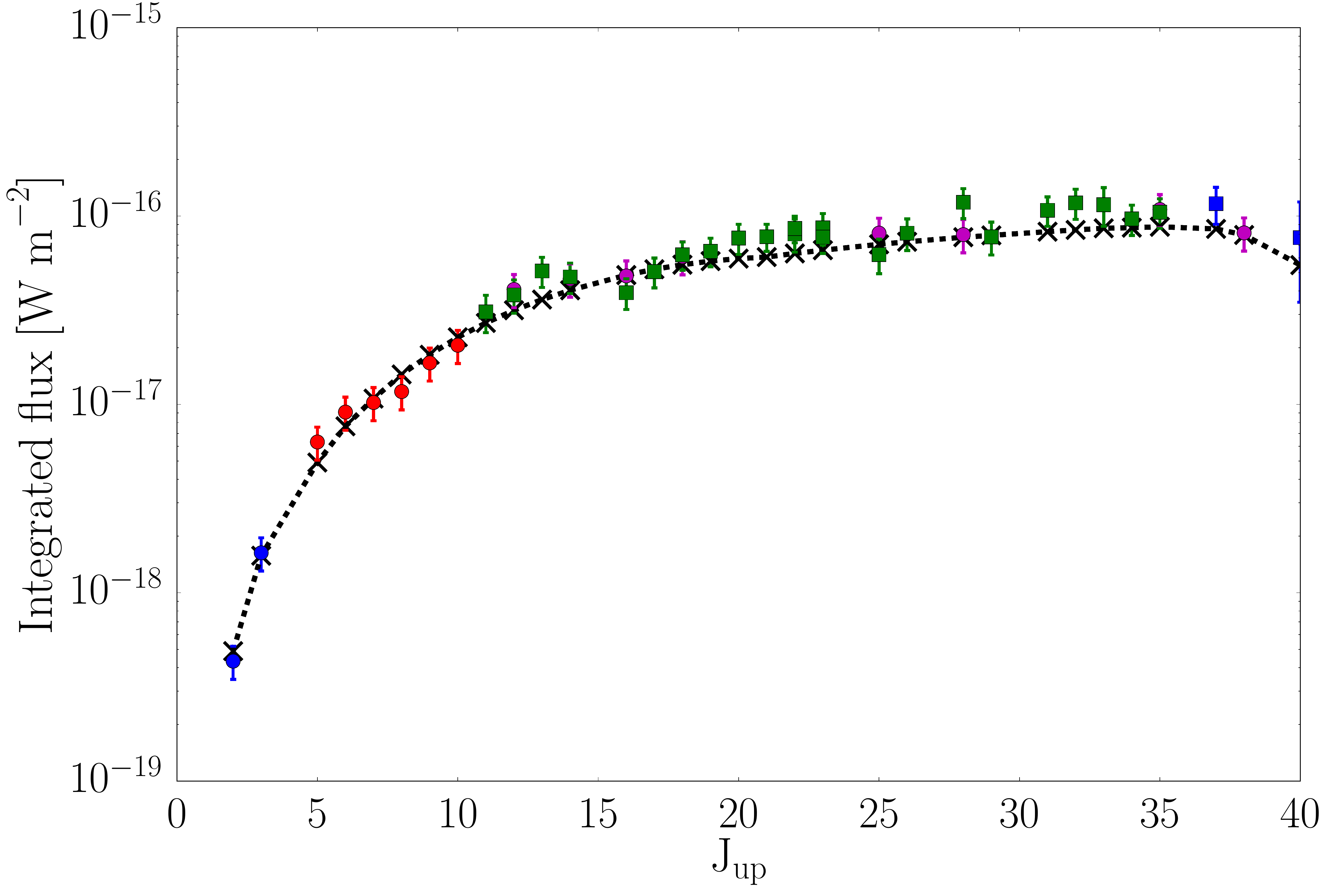}
\caption{Black: integrated line strengths of SiO modelled using the abundance profile characterised by $f_0 = 6.0 \times 10^{-5}$ relative to H$_2$ and $e$-folding radius $r_e = 4 \times 10^{13}$ cm with $R_{\rm decl} = 50\ R_*$. Overplotted are the integrated line strengths measured with APEX (red circles), SEST (blue circles), HIFI (purple circles), SPIRE (blue squares), and PACS (green squares). The errorbars for the resolved molecular lines correspond to an absolute error of 20\%.
}
\label{fig:Results-SiO-IntInt}
\end{figure}

\begin{figure}
\centering
\includegraphics[width=\columnwidth]{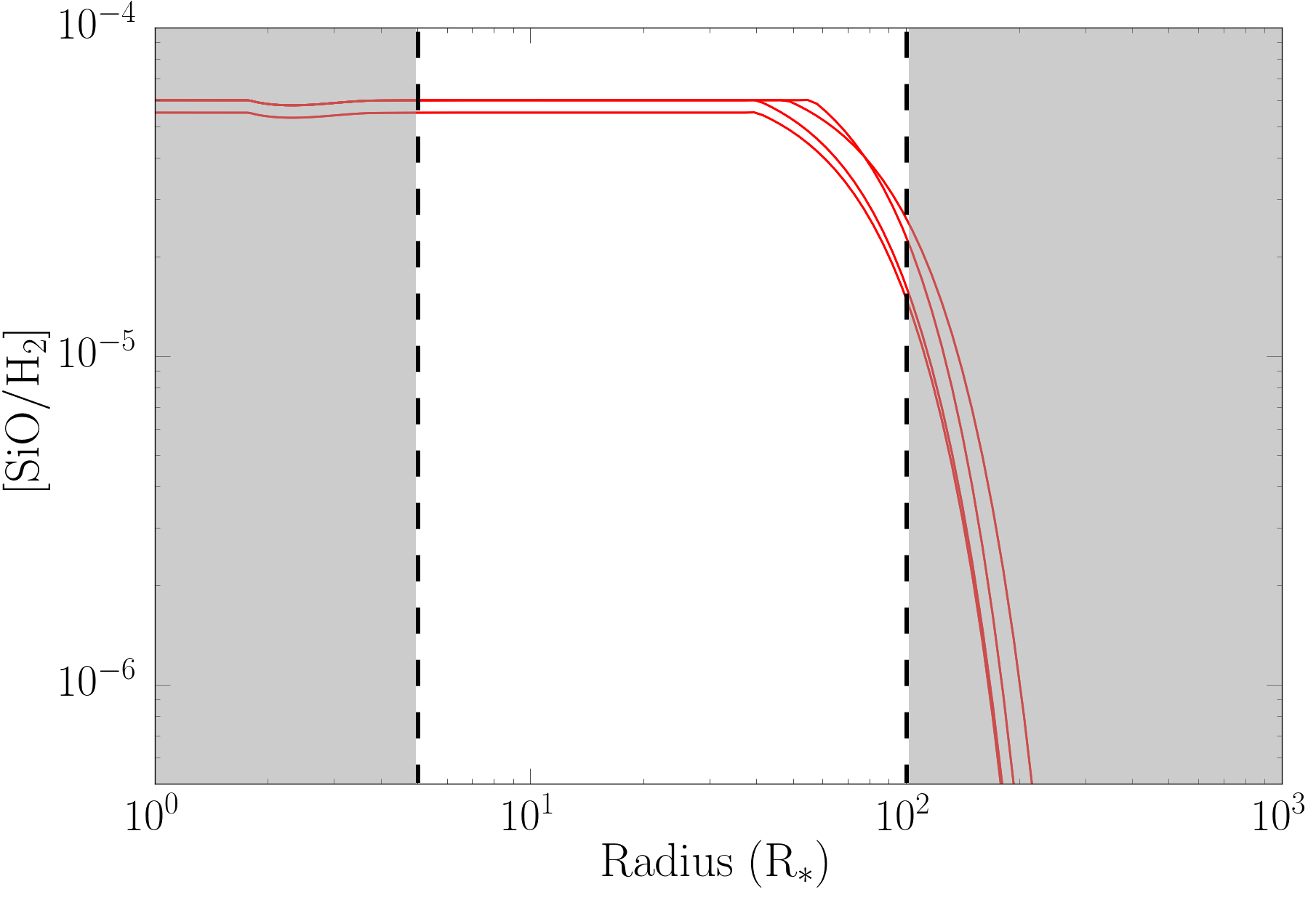}
\caption{The range in abundance profiles found for SiO throughout the CSE. The molecular transitions probe the region marked between dashed vertical lines.
}
\label{fig:Results-SiO-AbunProf}
\end{figure}

The SiO $J = 6-5$ and $5-4$ rotational transitions have been observed both by the APEX and SEST telescope. The integrated intensities of both measurements (see Table \ref{table:Data-Reso}) differ by 19\% and 20\% respectively, with the SEST observations having the lower integrated line strength. Because of this discrepancy and because the APEX observations were part of our APEX spectral scan (see Sect. \ref{subsect:Data-APEX}) yielding most of the SiO rotational transitions with $J_{\mathrm{up}} < 12$, we have excluded the SEST transitions when determining the SiO abundance profile. 
In App. \ref{sect:App:Luminosity}, we expand on the discrepancy in integrated intensities.

Fig. \ref{fig:Results-SiO-LPsGood} shows the 16 resolved rotational transitions in the ground vibrational state of SiO that were used to constrain the abundance profile.
All resolved molecular transitions with $J_{\mathrm{up}} > 16$ are noisy (see Sect. \ref{sect:Data}).
The molecular data probe a region between approximately 5 to 100 $R_*$.
We find that the initial abundance $f_0$ lies between $5.5 \times 10^{-5}$ and $6.0 \times 10^{-5}$ with respect to H$_2$, the $e$-folding radius between $3.5 \pm 0.5 \times 10^{13}$ cm  or $1.4 \pm 1.2\ R_* $, and the onset of the Gaussian decline $R_\mathrm{decl}$ between $60\ \pm 10\  R_*$ (see Table \ref{table:Results-AbunRange}).
The range in $f_0$ was explored using a stepsize of $0.5 \times 10^{-5}$, that of $r_e$ using a stepsize of $1 \times 10^{13}$ cm, and that of $R_\mathrm{decl}$ using a stepsize of 10 $R_*$.
Not all possible combinations result in good models, the specific combinations are listed in Table \ref{table:Results-AbunRange-SiO}.
The abundance profiles are shown in Fig. \ref{fig:Results-SiO-AbunProf} together with the region probed by the molecular data.
The results for the PACS and SPIRE data, along with the resolved data, are shown in Fig. \ref{fig:Results-SiO-IntInt}. 
Fig. \ref{fig:Results-SiO-LPsGood} and \ref{fig:Results-SiO-IntInt} show the results for one of the (equivalent) abundance profiles.
Henceforth, we use the abundance profile characterised by $f_0 = 6.0 \times 10^{-5}$ relative to H$_2$ and $e$-folding radius $r_e = 4 \times 10^{13}$ cm with $R_{\rm decl} = 50\ \mathrm{R}_*$ to compare our models with the data in all plots.
The modelled PACS spectra can be found in App. \ref{app:PACSSpectra}.

The goodness-of-fit of the resulting models was determined using the two criteria defined in Sect. \ref{subsect:Meth-GoodnessOfFit}.
The error on the integrated line strength was assumed to be 25\%, a small deviation from the commonly assumed 20\% (see Sect. \ref{subsect:Data-HIFI}). For the noisy lines, we have assumed an error of 35\%. Accepting these errors, all 16 rotational transitions satisfy both criteria.

An emission component is apparent in the blue wing of the rotational transitions, that is more prominent for higher $J$-values, as can be seen in Fig. \ref{fig:Results-SiO-LPsGood}.
The blue-wing feature is too prominent and inconsistently visible in the different line profiles to be caused by self-absorption only \citep{Morris1985,Schoenberg1988}.
Recently obtained ALMA data of R Dor (P.I. L. Decin) suggest that the feature is related to the morphology of the CSE of R Dor (Decin, priv. comm.). 
The feature has been included in our calculation of the integrated line strengths and in our goodness-of-fit analysis.
It is therefore likely that due to our assumption of spherical symmetry, we cannot reproduce the prominent blue-wing feature as an extra emission component is necessary in this velocity range.

\begin{table}
	\caption{
	Allowed parameter combinations of the abundance profiles of SiO. The parameters are described in Sect. \ref{subsect:Meth-GASTRoNOoM}.
	} 
	\centering
	%\resizebox{0.98\columnwidth}{!}{%
    \begin{tabular}{c c c }
    \hline \hline 
    \noalign{\smallskip}
$R_\mathrm{decl} $ [R$_*$]     &    $f_0$ [wrt H$_2$]      &    $r_e$  [cm]  \\     
\hline
\noalign{\smallskip}
50    &    $5.5 \times 10^{-5}$    &    $4 \times 10^{13}$    \\
50    &    $6.0 \times 10^{-5}$    &    $4 \times 10^{13}$    \\
%60    &    $6.0 \times 10^{-5}$    &    $3 \times 10^{13}$    \\
60    &    $6.0 \times 10^{-5}$    &    $4 \times 10^{13}$    \\
70    &    $6.0 \times 10^{-5}$    &    $3 \times 10^{13}$    \\
%\noalign{\smallskip}
    \hline
    \end{tabular}%
   %}
    \label{table:Results-AbunRange-SiO}    
\end{table}

%----------------------------------------------------------------------------------------------------------
\subsection{Abundance profile of HCN}			\label{subsect:Results-AbunProf-HCN}
%----------------------------------------------------------------------------------------------------------

\begin{figure*}[h!]
\centering
\includegraphics[width=1.\textwidth]{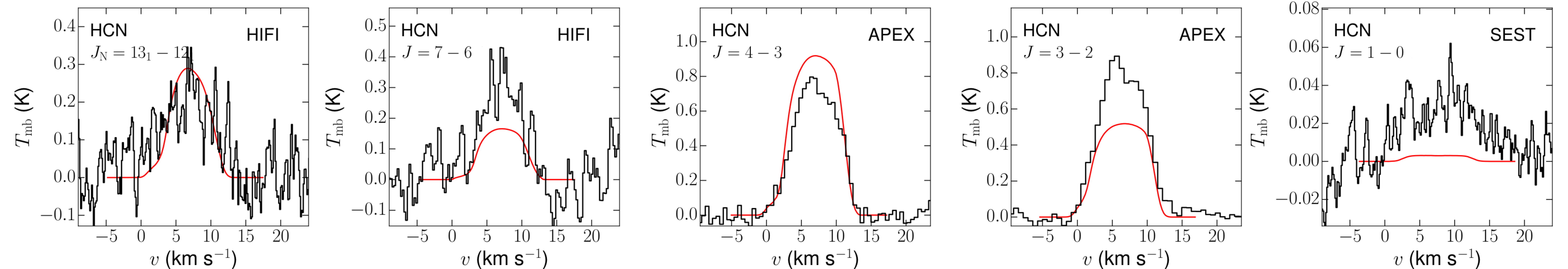}
\caption{Black: observed main-beam temperature as a function of velocity for 5 rotational transitions of HCN (see Table \ref{table:Data-Reso}). The telescope and rotational quantum number are indicated for each line.
Red: resulting line profiles using a Gaussian abundance profile characterised by $f_0 = 5.0 \times 10^{-7}$ with respect to H$_2$ and $r_\mathrm{e} = 1.8 \times 10^{15}$ cm. 
The $\upsilon_{\mathrm{LSR}}$ of R Dor is 7.5 km s$^{-1}$.
}
\label{fig:Results-HCN-LPs}
\end{figure*}

\begin{figure}
\centering
\includegraphics[width=0.98\columnwidth]{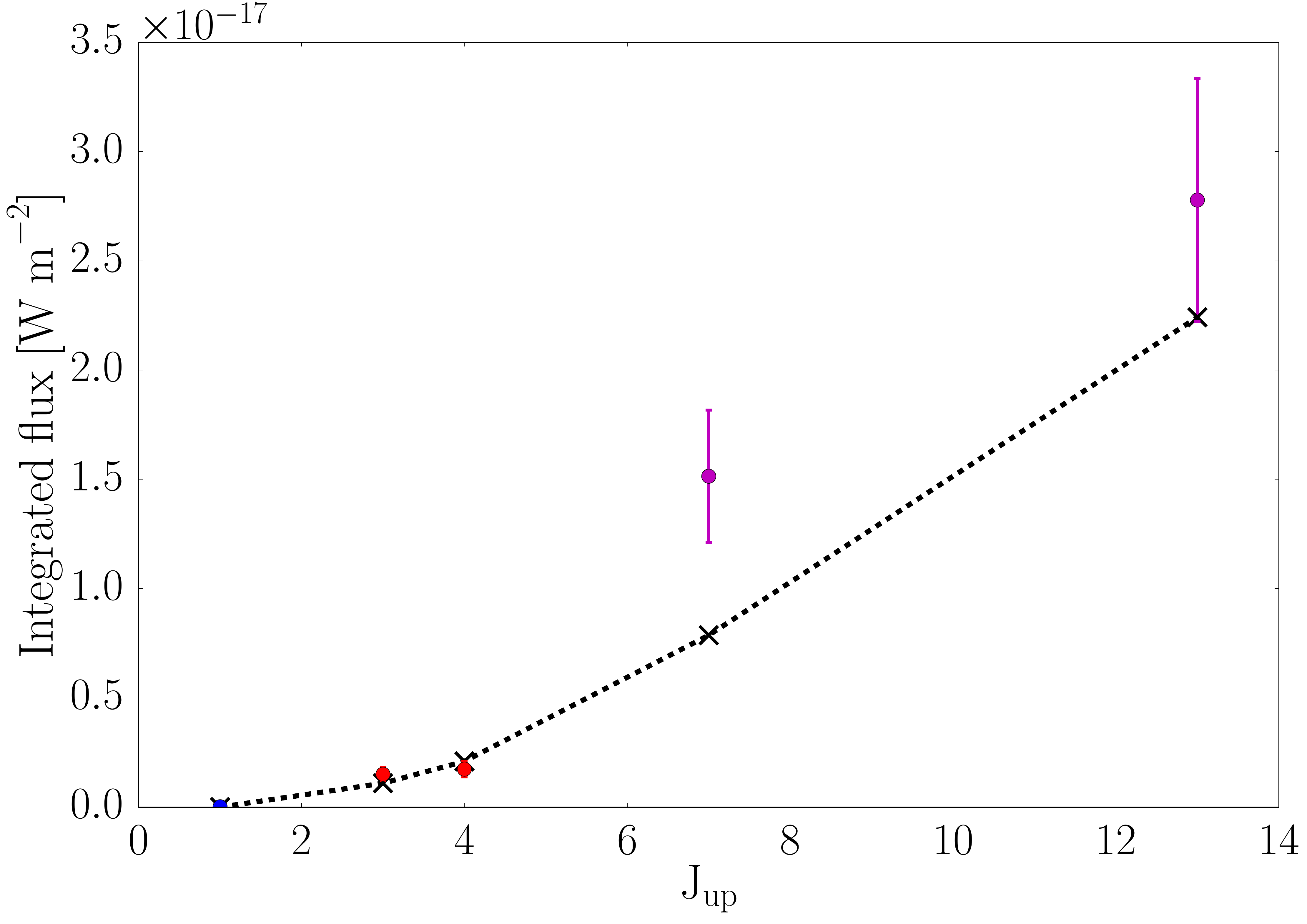}
\caption{Black: integrated line strengths of HCN modelled using a Gaussian abundance profile characterised by $f_0$ = $5.0\times 10^{-7}$ with respect to H$_2$  and $r_e = 1.8 \times 10^{15}$ cm. Overplotted are the integrated line strengths measured with APEX (red circles), SEST (blue circles), and HIFI (purple circles). The errorbars for the resolved molecular lines correspond to an absolute error of 20\%.
}
\label{fig:Results-HCN-IntInt}
\end{figure}

\begin{figure}
\centering
\includegraphics[width=\columnwidth]{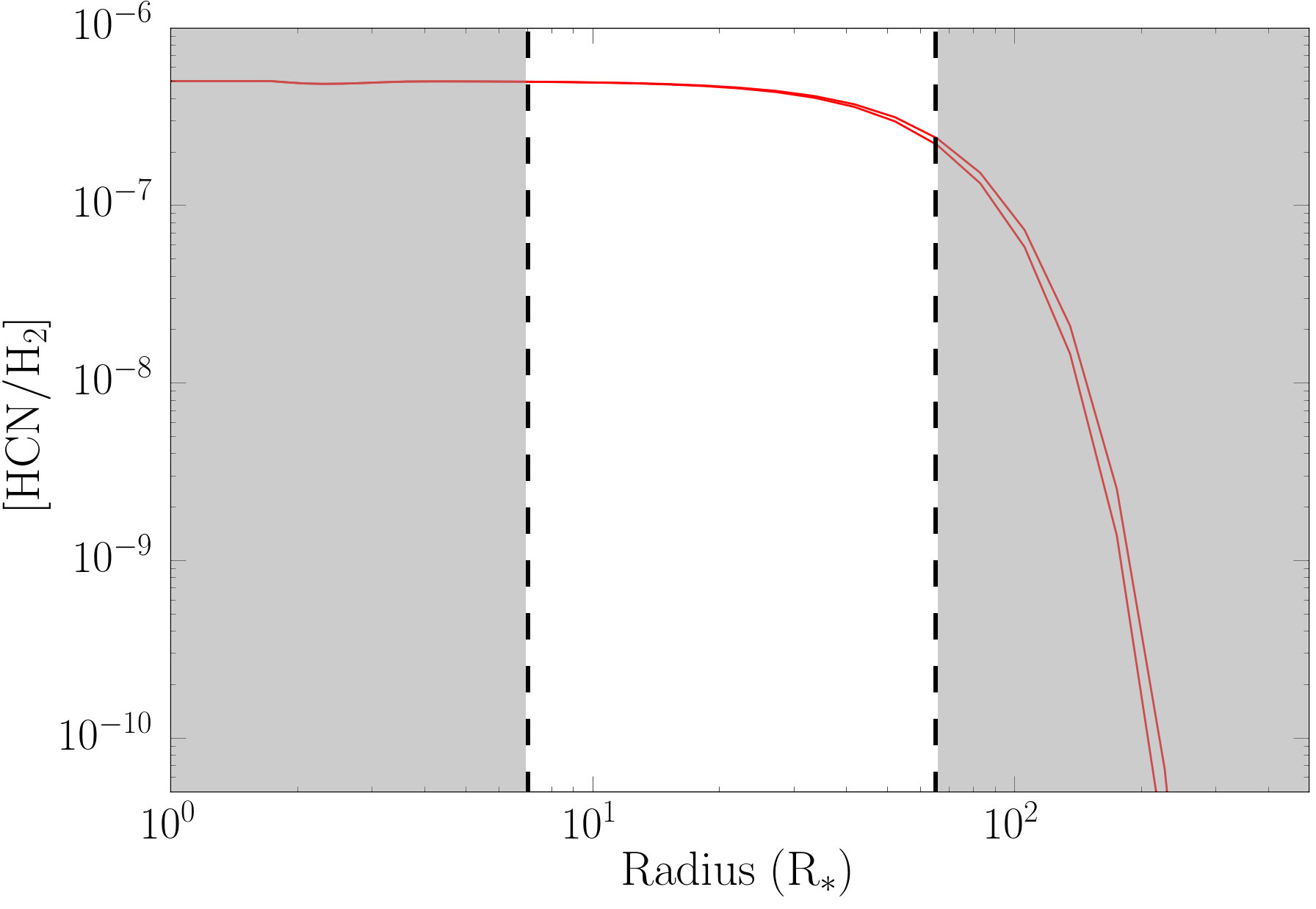}
\caption{The range in abundance profiles found for HCN throughout the CSE. The molecular transitions probe the region marked between dashed vertical lines.
}
\label{fig:Results-HCN-AbunProf}
\end{figure}

For HCN, we have used five resolved molecular lines in the ground vibrational state to determine the abundance profile, see Fig. \ref{fig:Results-HCN-LPs}.
The molecular data probe a region between approximately 7 and 40 $R_*$.
We find that the initial abundance $f_0$ lies between $5.0 \times 10^{-7}$ and $7.5 \times 10^{-7}$ with respect to H$_2$ and the $e$-folding radius $r_e$ lies between $1.35 \pm 0.25 \times 10^{15}$ cm or $0.54 \pm 0.10\ R_*$ (see Table \ref{table:Results-AbunRange}). 
The range in $f_0$ was explored using a stepsize of $0.5 \times 10^{-7}$, that of $r_e$ using a stepsize of $0.1 \times 10^{15}$ cm.
Not all possible combinations result in good models: larger values of $r_e$ correspond to smaller values of $f_0$.
The abundance profiles are shown in Fig. \ref{fig:Results-HCN-AbunProf} together with the region probed by the molecular data.
Fig. \ref{fig:Results-HCN-LPs} and \ref{fig:Results-HCN-IntInt} show the results for one of the (equivalent) abundance profiles.
Henceforth, we use the abundance profile characterised by $f_0 = 5.0 \times 10^{-7}$ with respect to H$_2$ and $r_\mathrm{e} = 1.8 \times 10^{15}$ cm to compare our models with the data in all plots.

The goodness-of-fit of the resulting models was determined following the two criteria in Sect. \ref{subsect:Meth-GoodnessOfFit}. 
We have deviated from the commonly assumed error on the integrated line strength of 20\% for all five molecular lines (see Sect. \ref{subsect:Data-HIFI}). 
For the $J=4-3$ and $3-2$ transitions, we have required that the integrated line strength lies within 30\% of its empirical value. This enables us to simultaneously fit both molecular lines.
The other three molecular lines all have a low SNR. For the $J = 13-12$ and $7-6$ transitions (SNR = 3.7 and 5.8 respectively), we require that the integrated line strength lies within 50\% of that of the observed strength.
The observation of the $J = 1-0$ transition has a SNR of 2.3 and is very weak, with a badly determined continuum. The transition is consistently underestimated in all models, which likely is caused by the neglect of masering in the models.
We therefore decided to exclude it from the analysis.  
Assuming these errors and excluding the $J = 1-0$ transition, all remaining 4 rotational transitions satisfy both goodness-of-fit criteria.

\begin{table}
	\caption{
	Parameter ranges of the abundance profiles of SiO and HCN for which fits to the molecular data are found (see Fig. \ref{fig:Results-SiO-AbunProf} and Fig. \ref{fig:Results-HCN-AbunProf}). The parameters are described in Sect. \ref{subsect:Meth-GASTRoNOoM}.
	} 
	\centering
	%\resizebox{0.98\columnwidth}{!}{%
    \begin{tabular}{c c c c }
    \hline \hline 
    \noalign{\smallskip}
Molecule        &    $R_\mathrm{decl} $ [R$_*$]     &    $f_0$ [wrt H$_2$]      &    $r_e$  [cm]  \\     
\hline
\noalign{\smallskip}
SiO    &  $50 - 70$      & $5.5 - 6.0  \times 10 ^{-5}$    &    $3.0 - 4.0 \times 10^{13}$ \\
%\noalign{\smallskip}
%HCN    &  0   & $5.0 - 7.5 \times 10^{-7}$    &    $1.1 - 1.6 \times 10^{15}$            \\
HCN    &  0   & $5.0  \times 10^{-7}$    &    $1.8 - 1.9 \times 10^{15}$            \\
%    \noalign{\smallskip}
    \hline
    \end{tabular}%
   %}
    \label{table:Results-AbunRange}    
\end{table}

%%%%%%%%%%%%%%%%%%%%%%%%%%%%%%%%%%%%%%%%%%%%%%%%%%%%%%%
\section{Discussion}					\label{sect:Discussion}
%%%%%%%%%%%%%%%%%%%%%%%%%%%%%%%%%%%%%%%%%%%%%%%%%%%%%%%

We discuss the range in abundance profiles found for SiO and HCN in Sect. \ref{subsect:Disc:SiO} and \ref{subsect:Disc:HCN}. 
We compare our results to those obtained previously for R Dor and IK Tau, the latter being a high-mass loss M-type AGB star.
These results were obtained using both retrieval methods and forward chemistry modelling. However, we cannot compare to forward chemistry results for R Dor, as, to the best of our knowledge, such analyses have not yet been done.

%----------------------------------------------------------------------------------------------------------
\subsection{Abundance stratification of SiO}                    \label{subsect:Disc:SiO}
%----------------------------------------------------------------------------------------------------------

Fig. \ref{fig:Results-SiO-AbunProf} shows the range in abundance profiles found for SiO together with the region in the stellar wind that is probed by the molecular data.
The abundance is constant up to the radius $R_{\rm decl}$, after which the abundance declines following a Guassian profile.

%The allowed range in $R_{\rm decl}$ (see Table \ref{table:Results-AbunRange}) is centered on the inner radius of the silicate envelope $R_\mathrm{Sil}$ at 60 R$_*$. The decline in abundance hence might be due to condensation onto dust grains. 
%Because of the low mass-loss rate, the decline in abundance could also be caused by photodissociation of SiO.
%Moreover, the location of $R_\mathrm{Sil}$ is dependent on the dust mass-loss rate (see Sect. \ref{subsubsect:Results-EnvModel-Dust}). To lift this degeneracy, the location of the dust shell needs to be properly imaged.
%We are therefore unable to distinguish between the two possible mechanisms.

The deduced range in $R_{\rm decl}$ (see Table \ref{table:Results-AbunRange}) coincides with the inner radius of the silicate envelope $R_\mathrm{Sil}$ at 60 R$_*$. The decline in abundance hence might be due to condensation onto dust grains. 
However, the location of $R_\mathrm{Sil}$ is degenerate in the dust mass-loss rate and the metallic iron content (see Sect. \ref{subsubsect:Results-EnvModel-Dust}). 
%To lift this degeneracy, the location of the dust shell needs to be properly imaged. 
Photodissociation caused by interstellar UV photons is an alternative mechanism to explain the decrease in SiO.
%As discussed in Sect. \ref{subsubsect:SiO-comparison}, the onset of photodissociation shifts inward with decreasing mass-loss rate.
The decline in abundance could also be caused by photodissociation of SiO, since the onset of photodissociation shifts inward with decreasing mass-loss rate (see Sect. \ref{subsubsect:SiO-comparison}). 
A way to differentiate between both mechanisms is by imaging the location of the silicate dust shell. No proper VLTI/MIDI data are available for R Dor to answer this question \citep{Decin2017}.
We are therefore unable to distinguish between the two possible mechanisms.

%--.--.--.--.--.--.--.--.--.--.--.--.--.--.--.--.--.--.--.--.--.--.--.--.--.--.--.--.--.--.--.--.--.--.--.--.--.--.--
\subsubsection{Comparison to previous results}			\label{subsubsect:SiO-comparison}
%--.--.--.--.--.--.--.--.--.--.--.--.--.--.--.--.--.--.--.--.--.--.--.--.--.--.--.--.--.--.--.--.--.--.--.--.--.--.--
%--.--.--.--.--.--.--.--.--.--.--.--.--.--.--.--.--.--.--.--.--.--.--.--.--.--.--.--.--.--.--.--.--.--.--.--.--.--.--
\paragraph{R Dor}
%--.--.--.--.--.--.--.--.--.--.--.--.--.--.--.--.--.--.--.--.--.--.--.--.--.--.--.--.--.--.--.--.--.--.--.--.--.--.--
The abundance profile of SiO has previously been determined by \citet{GonzalezDelgado2003}, \citet{Schoier2004}, and \citet{KhouriPhD}.
\citet{GonzalezDelgado2003} used the $J = 2-1$, $3-2$, $5-4$, and $6-5$ transitions, which are all included in this paper (see Table \ref{table:Data-Reso}). They found a Gaussian abundance profile characterised by an initial abundance $f_0$ = $5 \times 10^{-6}$ and $e$-folding radius $r_\mathrm{e} = 3.3 \times 10^{15}$. 
We find a larger initial abundance, which might be due to a combination of our inclusion of higher-$J$ molecular lines and differences in the modelling of the envelope, e.g. the velocity profile.

\citet{Schoier2004} supplemented the \citet{GonzalezDelgado2003} data set by $J = 1-0$ ground and first excited vibrational level observations obtained with the Australia Telescope Compact Array. They found that two Gaussian profiles fit the data better: a high-density, compact component (C) together with a low-density, more extended component (E). They are characterised by $f_\mathrm{C}$ = $4 \times 10^{-5}$ and $r_\mathrm{e,C} = 1.2 \times 10^{15}$ cm, and $f_\mathrm{E}$ = $3 \times 10^{-6}$ and $r_\mathrm{e,E} = 3.3 \times 10^{15}$ cm respectively.
We find no indications for a two-component abundance profile.

The abundance profile derived in \citet{KhouriPhD} is based on that of \citet{Schoier2004}. The same $e$-folding radius was adopted, but with a higher initial abundance $f_0$(SiO) = $6 \times 10^{-5}$. The profile was determined using HIFI and SPIRE data, see Tables \ref{table:Data-Reso} and \ref{table:Data-Unreso-SPIRE}.
We find a similar initial abundance, but a smaller $e$-folding radius. This is likely due to our inclusion of lower-$J$ molecular lines.

%--.--.--.--.--.--.--.--.--.--.--.--.--.--.--.--.--.--.--.--.--.--.--.--.--.--.--.--.--.--.--.--.--.--.--.--.--.--.--
\paragraph{IK Tau}
%--.--.--.--.--.--.--.--.--.--.--.--.--.--.--.--.--.--.--.--.--.--.--.--.--.--.--.--.--.--.--.--.--.--.--.--.--.--.--

\citet{Decin2010} constrained the SiO abundance of the stellar outflow of IK Tau and found an initial abundance of $1.6 \times 10^{-5}$, which decreases with a factor of $\sim$ 40 at approximately 180 $R_*$. 
We do not find such a decline in the SiO abundance profile for R Dor. 
The difference in initial abundance between R Dor and IK Tau of a factor of 3.5 might be connected to differences in the dust condensation physics.
\citet{Hofner2016} described condensation of silicates onto Al$_2$O$_3$ seed particles. This process may take place in the GBDS, but its outcome may differ between R Dor and IK Tau due to their different pulsation characteristics, such as period and shock strength.
The photodissociation of SiO takes place far out in the wind, around a few thousand stellar radii.
For R Dor, the destruction of SiO occurs much closer to the star. This is likely due to its lower mass-loss rate, as interstellar UV photons are able to penetrate regions closer to the central star.

\citet{Bujarrabal1989} and \citet{Sahai1993} determined the initial SiO abundance for a sample of O-rich AGB stars, including IK Tau (but not R Dor). They deduced a value of $5 \times 10^{-5}$ and a lower limit of $1.0 \times 10^{-5}$ respectively. 
%These higher abundances correspond to our initial SiO abundance in R Dor. 
Our deduced abundance corresponds well to that of \citet{Sahai1993} and is compatible with the result of \citet{Bujarrabal1989}.

\citet{Lucas1992} have determined the extent of the SiO envelope for IK Tau and other O-rich AGB stars. 
They found that the extent of SiO ranges from $0.6 \times 10^{15}$ to $7.5 \times 10^{15}$ cm. Our deduced range for $R_\mathrm{decl}$ lies within this range.

The abundance of SiO in the stellar outflow of IK Tau was also modelled using forward chemistry models.
\citet{Gobrecht2016} have modelled the inner wind region up to the dust formation zone and found SiO abundances that corresponded well with the observations of \citet{Decin2010}.
\citet{Li2016} have modelled the chemistry in the outer wind, using a gas-phas only chemical network. They assumed SiO to be one of the parent species, i.e. one of the species present immediately beyond the dust formation zone. The initial abundance in the chemical network was assumed to be that measured by \citet{Decin2010}. \citet{Li2016} found a decrease of approximately an order of magnitude at $\sim$ 1000 $R_*$, mainly caused by photodissociation.
We find that the decline in abundance for R Dor starts at more than an order of magnitude in radius closer to the central star. 
However, the onset of photodissociation shifts inwards with decreasing mass-loss rate, approximately one order of magnitude in radius per order of magnitude in mass-loss rate \citep{Mamon1988}.
%as the visual extinction throughout the wind scales with mass-loss rate.
The mass-loss rates of R Dor and IK Tau defer by a factor of $\sim$ 60.
Therefore, the onset of photodissociation in R Dor can correspond to the range found in $R_\mathrm{decl}$.

%We find that our range in abundance profiles shows a similar behaviour as that modelled by \citet{Li2016}, supporting the hypothesis that photodissociation is the main destruction mechanism.

%----------------------------------------------------------------------------------------------------------
\subsection{Abundance stratification of HCN}                    \label{subsect:Disc:HCN}
%----------------------------------------------------------------------------------------------------------

Fig. \ref{fig:Results-HCN-AbunProf} shows the abundance profiles found for HCN together with the region probed by the molecular data. 
The abundance profile is a Gaussian profile centred on the star.
We find that a range in parameters produce good fits to the data, see Table \ref{table:Results-AbunRange}.
The decline in HCN abundance is most likely due to photodissociation, as it takes place in the outer wind and HCN is not a refractory species.

%--.--.--.--.--.--.--.--.--.--.--.--.--.--.--.--.--.--.--.--.--.--.--.--.--.--.--.--.--.--.--.--.--.--.--.--.--.--.--
\subsubsection{Comparison to previous results}		\label{subsubsect:HCN-comparison}
%--.--.--.--.--.--.--.--.--.--.--.--.--.--.--.--.--.--.--.--.--.--.--.--.--.--.--.--.--.--.--.--.--.--.--.--.--.--.--
%--.--.--.--.--.--.--.--.--.--.--.--.--.--.--.--.--.--.--.--.--.--.--.--.--.--.--.--.--.--.--.--.--.--.--.--.--.--.--
\paragraph{R Dor}
%--.--.--.--.--.--.--.--.--.--.--.--.--.--.--.--.--.--.--.--.--.--.--.--.--.--.--.--.--.--.--.--.--.--.--.--.--.--.--

\citet{Schoier2013} have measured the HCN abundance using the $J=1-0$, $3-2$, and $4-3$ rotational transitions in the ground vibrational state. Their measurement of the $J=1-0$ transition is included in this work.
\citet{Schoier2013} have used a Gaussian abundance profile, characterised by an initial abundance $f_0$(HCN) = $1.0  \times 10^{-7}$ with respect to H$_2$ and an $e$-folding radius $r_\mathrm{e} = 5.1 \times 10^{15}$ cm. 

We find Gaussian abundance profiles with a larger initial HCN abundance than \citet{Schoier2013}, a difference of roughly a factor of 6. 
This discrepancy may be explained by the different data sets used, since we have included higher-$J$ transitions that are excited at higher temperatures deeper inside the outflow. Alternatively, it could be due to different modelling assumptions, e.g. the velocity profile. 
%Nevertheless, the initial abundances lie within an order of magnitude of each other.
The onset of photodissociation starts closer to the star compared to \citet{Schoier2013}, a difference of roughly a factor of 4. 
In both models, the onset is located in the outer wind.

%--.--.--.--.--.--.--.--.--.--.--.--.--.--.--.--.--.--.--.--.--.--.--.--.--.--.--.--.--.--.--.--.--.--.--.--.--.--.--
\paragraph{IK Tau}
%--.--.--.--.--.--.--.--.--.--.--.--.--.--.--.--.--.--.--.--.--.--.--.--.--.--.--.--.--.--.--.--.--.--.--.--.--.--.--

\citet{Decin2010} have measured the HCN abundance of the outflow of IK Tau and found an initial abundance of $4.4 \times 10^{-7}$. The start of the decline in abundance is located at 500 $R_*$. 
The initial abundance found for R Dor is hence slightly higher by a factor of 1.1 than that found for IK Tau, and the decline in abundance occurs closer to the star.
The earlier onset of photodissociation of HCN in R Dor is likely due to its low mass-loss rate.

In the forward chemistry modelling of the inner wind region by \citet{Gobrecht2016}, which accounts for the effects of periodic shocks, the HCN abundance was found to be consistent with the observations of \citet{Decin2010}.
\citet{Li2016} have modelled the chemistry in the outer wind and have assumed HCN to be one of the parent species. The inital abundance in the chemical network was assumed to be that measured by \citet{Decin2010}. They found that the abundance declines at $\sim$ 1000 $R_*$, hence in the outer wind.
We find that our range in abundance profiles shows a similar behaviour.
This supports the hypothesis that the decline in abundance is due to photodissociation, as this is the main destructive chemical reaction in the outer wind found by \citet{Li2016}.

The initial HCN abundance in both R Dor and IK Tau is about five orders of magnitude larger than predicted by thermal equilibrium chemistry, which indicates the importance of non-equilibrium chemistry in the inner wind. Moreover, as the initial HCN abundance of R Dor and IK Tau lie within a factor of 2 of each other, quite similar physical and chemical conditions are likely present in their inner winds independent of mass-loss rate.
The factor of 2 difference might be due to the lower mass-loss rate of R Dor or different pulsation characteristics.

%%%%%%%%%%%%%%%%%%%%%%%%%%%%%%%%%%%%%%%%%%%%%%%%%%%%%%%
\section{Conclusions}					\label{sect:Conclusions}
%%%%%%%%%%%%%%%%%%%%%%%%%%%%%%%%%%%%%%%%%%%%%%%%%%%%%%%

We have constrained the radial abundance profile of SiO and HCN in the stellar wind of the low mass-loss M-type AGB star R Dor using a non-LTE extended atmosphere model that accounts for the interaction between gas and dust. 
%The atmosphere model is based on the work of \citet{KhouriPhD}.
Our analysis used the integrated line intensity and line profile shapes of a set of lines obtained with both ground- and space-based instruments.  
For SiO, we have used 42 molecular transitions, of which 16 are spectrally resolved. The molecular data are sensitive to the region between approximately 5 and 100 $R_*$. 
The initial abundance of SiO lies between $5.5 \times 10^{-5}$ and $6.0 \times 10^{-5}$ with respect to H$_2$. The abundance profile is constant up to $60\ \pm 10 \ R_*$, after which it follows a Gaussian profile with an $e$-folding radius between $3.5 \pm 0.5 \times 10^{13}$ cm.
The decline in SiO abundance near $\sim$ 60 R$_*$ might be due to condensation onto dust grains or photodissociation. We are unable to unambiguously determine the mechanism at work.
For HCN, we have used 5 resolved molecular transitions. The molecular data are sensitive to the region between approximately 7 and 40 $R_*$.
The initial abundance of HCN lies between $5.0 \times 10^{-7}$ and $7.5 \times 10^{-7}$ with respect to H$_2$. The abundance profile is a Gaussian profile centred on the star with an $e$-folding radius $r_e$ between $1.35 \pm 0.25 \times 10^{15}$ cm. 

The intial abundances found are larger than those reported in previous studies (with the exception of the study by \citet{KhouriPhD} for SiO). This might be due to our inclusion of higher-$J$ molecular lines. 
Compared to results for IK Tau, a high mass-loss rate M-type AGB star, the onset of photodissociation occurs closer to the star for both SiO and HCN. This can be explained by their difference in mass-loss rate. 
The R Dor initial abundance of SiO is also larger than those found for IK Tau. 
This might be due to different pulsation characteristics of the central stars and/or a difference in dust condensation physics.
It is not connected to a differing sensitivity to non-equilibrium chemistry in the inner wind, as this molecule is unaffected by such processes \citep{Cherchneff2006}.
For both stars, we find roughly similar HCN abundances, though both are five orders of magnitude larger than the abundance predicted by thermal equilibrium chemistry. This suggests similar physical and chemical conditions in the inner wind, independent of mass-loss rate.

The interaction between micro-scale chemistry and macro-scale dynamics in the stellar outflow can be disentangled by comparing retrieved abundances to those predicted by forward chemistry models.
The abundance profiles for SiO and HCN we have retrieved are complementary to those previously determined for other molecules in the outflow of R Dor, e.g. SO and SO$_2$ \citep{Danilovich2016} and H$_2$O \citep{Maercker2016}. 
Moreover, they are complementary to those determined by \citet{Decin2010} for a variety of molecules in the outflow of IK Tau. 
Together with recently obtained ALMA data (PI L. Decin) of R Dor and IK Tau, which will unveil the molecular abundances in their dust formation regions, our results furthermore fit into the larger framework of unravelling the intricacies of dust formation in M-type AGB stars.

%-------------------------------------------------------------------
\begin{acknowledgements}
MVdS and LD acknowledge support from the Research Council of the KU Leuven under grant number GOA/2013/012. 
LD acknowledges support from the ERC consolidator grant 646758 AEROSOL and the FWO Research Project grant G024112N.
We acknowledge the variable star observations from the AAVSO International Database contributed by observers worldwide and used in this paper.
This paper has made use of the VizieR catalogue access tool, CDS, Strasbourg, France.
The authors thank the anonymous referee for their constructive comments.
\end{acknowledgements}
%-------------------------------------------------------------------

\bibliographystyle{aa}
\bibliography{rdor}

\begin{thebibliography}{75}
\expandafter\ifx\csname natexlab\endcsname\relax\def\natexlab#1{#1}\fi

\bibitem[{{Ag{\'u}ndez} \& {Cernicharo}(2006)}]{Agundez2006}
{Ag{\'u}ndez}, M. \& {Cernicharo}, J. 2006, \apj, 650, 374

\bibitem[{Bardeau \& Pety(2006)}]{Bardeau2006}
Bardeau, S. \& Pety, J. 2006, CLASS: Continuum and Line Analysis Single-Dish
  Software, \url{http://www.iram.fr/IRAMFR/GILDAS}

\bibitem[{{Bedding} {et~al.}(1998){Bedding}, {Zijlstra}, {Jones}, \&
  {Foster}}]{Bedding1998}
{Bedding}, T.~R., {Zijlstra}, A.~A., {Jones}, A., \& {Foster}, G. 1998, \mnras,
  301, 1073

\bibitem[{{Bujarrabal} {et~al.}(1989){Bujarrabal}, {Gomez-Gonzalez}, \&
  {Planesas}}]{Bujarrabal1989}
{Bujarrabal}, V., {Gomez-Gonzalez}, J., \& {Planesas}, P. 1989, \aap, 219, 256

\bibitem[{{Cernicharo} {et~al.}(2014){Cernicharo}, {Teyssier},
  {Quintana-Lacaci}, {Daniel}, {Ag{\'u}ndez}, {Velilla-Prieto}, {Decin},
  {Gu{\'e}lin}, {Encrenaz}, {Garc{\'{\i}}a-Lario}, {de Beck}, {Barlow},
  {Groenewegen}, {Neufeld}, \& {Pearson}}]{Cernicharo2014}
{Cernicharo}, J., {Teyssier}, D., {Quintana-Lacaci}, G., {et~al.} 2014, \apjl,
  796, L21

\bibitem[{{Cherchneff}(2006)}]{Cherchneff2006}
{Cherchneff}, I. 2006, \aap, 456, 1001

\bibitem[{{Cutri} {et~al.}(2003){Cutri}, {Skrutskie}, {van Dyk}, {Beichman},
  {Carpenter}, {Chester}, {Cambresy}, {Evans}, {Fowler}, {Gizis}, {Howard},
  {Huchra}, {Jarrett}, {Kopan}, {Kirkpatrick}, {Light}, {Marsh}, {McCallon},
  {Schneider}, {Stiening}, {Sykes}, {Weinberg}, {Wheaton}, {Wheelock}, \&
  {Zacarias}}]{2003yCat.2246....0C}
{Cutri}, R.~M., {Skrutskie}, M.~F., {van Dyk}, S., {et~al.} 2003, VizieR Online
  Data Catalog, 2246

\bibitem[{{Cutri} {et~al.}(2012){Cutri}, {Skrutskie}, {van Dyk}, {Beichman},
  {Carpenter}, {Chester}, {Cambresy}, {Evans}, {Fowler}, {Gizis}, {Howard},
  {Huchra}, {Jarrett}, {Kopan}, {Kirkpatrick}, {Light}, {Marsh}, {McCallon},
  {Schneider}, {Stiening}, {Sykes}, {Weinberg}, {Wheaton}, {Wheelock}, \&
  {Zacharias}}]{2012yCat.2281....0C}
{Cutri}, R.~M., {Skrutskie}, M.~F., {van Dyk}, S., {et~al.} 2012, VizieR Online
  Data Catalog, 2281

\bibitem[{{Daniel} {et~al.}(2012){Daniel}, {Ag{\'u}ndez}, {Cernicharo}, {De
  Beck}, {Lombaert}, {Decin}, {Kahane}, {Gu{\'e}lin}, \&
  {M{\"u}ller}}]{Daniel2012}
{Daniel}, F., {Ag{\'u}ndez}, M., {Cernicharo}, J., {et~al.} 2012, \aap, 542,
  A37

\bibitem[{{Danilovich} {et~al.}(2016){Danilovich}, {De Beck}, {Black},
  {Olofsson}, \& {Justtanont}}]{Danilovich2016}
{Danilovich}, T., {De Beck}, E., {Black}, J.~H., {Olofsson}, H., \&
  {Justtanont}, K. 2016, \aap, 588, A119

\bibitem[{{De Beck} {et~al.}(2010){De Beck}, {Decin}, {de Koter}, {Justtanont},
  {Verhoelst}, {Kemper}, \& {Menten}}]{DeBeck2010}
{De Beck}, E., {Decin}, L., {de Koter}, A., {et~al.} 2010, \aap, 523, A18

\bibitem[{{De Beck} {et~al.}(2012){De Beck}, {Lombaert}, {Ag{\'u}ndez},
  {Daniel}, {Decin}, {Cernicharo}, {M{\"u}ller}, {Min}, {Royer},
  {Vandenbussche}, {de Koter}, {Waters}, {Groenewegen}, {Barlow}, {Gu{\'e}lin},
  {Kahane}, {Pearson}, {Encrenaz}, {Szczerba}, \& {Schmidt}}]{DeBeck2012}
{De Beck}, E., {Lombaert}, R., {Ag{\'u}ndez}, M., {et~al.} 2012, \aap, 539,
  A108

\bibitem[{{de Graauw} {et~al.}(1996){de Graauw}, {Haser}, {Beintema},
  {Roelfsema}, {van Agthoven}, {Barl}, {Bauer}, {Bekenkamp}, {Boonstra},
  {Boxhoorn}, {Cote}, {de Groene}, {van Dijkhuizen}, {Drapatz}, {Evers},
  {Feuchtgruber}, {Frericks}, {Genzel}, {Haerendel}, {Heras}, {van der Hucht},
  {van der Hulst}, {Huygen}, {Jacobs}, {Jakob}, {Kamperman}, {Katterloher},
  {Kester}, {Kunze}, {Kussendrager}, {Lahuis}, {Lamers}, {Leech}, {van der
  Lei}, {van der Linden}, {Luinge}, {Lutz}, {Melzner}, {Morris}, {van Nguyen},
  {Ploeger}, {Price}, {Salama}, {Schaeidt}, {Sijm}, {Smoorenburg}, {Spakman},
  {Spoon}, {Steinmayer}, {Stoecker}, {Valentijn}, {Vandenbussche}, {Visser},
  {Waelkens}, {Waters}, {Wensink}, {Wesselius}, {Wiezorrek}, {Wieprecht},
  {Wijnbergen}, {Wildeman}, \& {Young}}]{DeGraauw1996}
{de Graauw}, T., {Haser}, L.~N., {Beintema}, D.~A., {et~al.} 1996, \aap, 315,
  L49

\bibitem[{{de Graauw} {et~al.}(2010){de Graauw}, {Helmich}, {Phillips},
  {Stutzki}, {Caux}, {Whyborn}, {Dieleman}, {Roelfsema}, {Aarts}, {Assendorp},
  {Bachiller}, {Baechtold}, {Barcia}, {Beintema}, {Belitsky}, {Benz}, {Bieber},
  {Boogert}, {Borys}, {Bumble}, {Ca{\"i}s}, {Caris}, {Cerulli-Irelli},
  {Chattopadhyay}, {Cherednichenko}, {Ciechanowicz}, {Coeur-Joly}, {Comito},
  {Cros}, {de Jonge}, {de Lange}, {Delforges}, {Delorme}, {den Boggende},
  {Desbat}, {Diez-Gonz{\'a}lez}, {di Giorgio}, {Dubbeldam}, {Edwards},
  {Eggens}, {Erickson}, {Evers}, {Fich}, {Finn}, {Franke}, {Gaier}, {Gal},
  {Gao}, {Gallego}, {Gauffre}, {Gill}, {Glenz}, {Golstein}, {Goulooze},
  {Gunsing}, {G{\"u}sten}, {Hartogh}, {Hatch}, {Higgins}, {Honingh}, {Huisman},
  {Jackson}, {Jacobs}, {Jacobs}, {Jarchow}, {Javadi}, {Jellema}, {Justen},
  {Karpov}, {Kasemann}, {Kawamura}, {Keizer}, {Kester}, {Klapwijk}, {Klein},
  {Kollberg}, {Kooi}, {Kooiman}, {Kopf}, {Krause}, {Krieg}, {Kramer},
  {Kruizenga}, {Kuhn}, {Laauwen}, {Lai}, {Larsson}, {Leduc}, {Leinz}, {Lin},
  {Liseau}, {Liu}, {Loose}, {L{\'o}pez-Fernandez}, {Lord}, {Luinge}, {Marston},
  {Mart{\'{\i}}n-Pintado}, {Maestrini}, {Maiwald}, {McCoey}, {Mehdi}, {Megej},
  {Melchior}, {Meinsma}, {Merkel}, {Michalska}, {Monstein}, {Moratschke},
  {Morris}, {Muller}, {Murphy}, {Naber}, {Natale}, {Nowosielski}, {Nuzzolo},
  {Olberg}, {Olbrich}, {Orfei}, {Orleanski}, {Ossenkopf}, {Peacock}, {Pearson},
  {Peron}, {Phillip-May}, {Piazzo}, {Planesas}, {Rataj}, {Ravera}, {Risacher},
  {Salez}, {Samoska}, {Saraceno}, {Schieder}, {Schlecht}, {Schl{\"o}der},
  {Schm{\"u}lling}, {Schultz}, {Schuster}, {Siebertz}, {Smit}, {Szczerba},
  {Shipman}, {Steinmetz}, {Stern}, {Stokroos}, {Teipen}, {Teyssier}, {Tils},
  {Trappe}, {van Baaren}, {van Leeuwen}, {van de Stadt}, {Visser}, {Wildeman},
  {Wafelbakker}, {Ward}, {Wesselius}, {Wild}, {Wulff}, {Wunsch}, {Tielens},
  {Zaal}, {Zirath}, {Zmuidzinas}, \& {Zwart}}]{DeGraauw2010}
{de Graauw}, T., {Helmich}, F.~P., {Phillips}, T.~G., {et~al.} 2010, \aap, 518,
  L6

\bibitem[{{De Nutte} {et~al.}(2017){De Nutte}, {Decin}, {Olofsson}, {Lombaert},
  {de Koter}, {Karakas}, {Milam}, {Ramstedt}, {Stancliffe}, {Homan}, \& {Van de
  Sande}}]{DeNutte2016}
{De Nutte}, R., {Decin}, L., {Olofsson}, H., {et~al.} 2017, \aap, 600, A71

\bibitem[{{Decin} {et~al.}(2010){Decin}, {De Beck}, {Br{\"u}nken},
  {M{\"u}ller}, {Menten}, {Kim}, {Willacy}, {de Koter}, \&
  {Wyrowski}}]{Decin2010}
{Decin}, L., {De Beck}, E., {Br{\"u}nken}, S., {et~al.} 2010, \aap, 516, A69

\bibitem[{{Decin} {et~al.}(2006){Decin}, {Hony}, {de Koter}, {Justtanont},
  {Tielens}, \& {Waters}}]{Decin2006}
{Decin}, L., {Hony}, S., {de Koter}, A., {et~al.} 2006, \aap, 456, 549

\bibitem[{{Decin} {et~al.}(2007){Decin}, {Hony}, {de Koter}, {Molenberghs},
  {Dehaes}, \& {Markwick-Kemper}}]{Decin2007}
{Decin}, L., {Hony}, S., {de Koter}, A., {et~al.} 2007, \aap, 475, 233

\bibitem[{{Decin} {et~al.}(2017){Decin}, {Richards}, {Waters}, {Danilovich},
  {Gobrecht}, {Khouri}, {Homan}, {Bakker}, {Van de Sande}, {Nuth}, \& {De
  Beck}}]{Decin2017}
{Decin}, L., {Richards}, A.~M.~S., {Waters}, L.~B.~F.~M., {et~al.} 2017, ArXiv
  e-prints [\eprint[arXiv]{1704.05237}]

\bibitem[{{Deguchi} \& {Uyemura}(1984)}]{Deguchi1984}
{Deguchi}, S. \& {Uyemura}, M. 1984, \apj, 285, 153

\bibitem[{{Duari} {et~al.}(1999){Duari}, {Cherchneff}, \&
  {Willacy}}]{Duari1999}
{Duari}, D., {Cherchneff}, I., \& {Willacy}, K. 1999, \aap, 341, L47

\bibitem[{{Ducati}(2002)}]{2002yCat.2237....0D}
{Ducati}, J.~R. 2002, VizieR Online Data Catalog, 2237

\bibitem[{{Flower}(1975)}]{Flower1975}
{Flower}, P.~J. 1975, \aap, 41, 391

\bibitem[{{Gail} \& {Sedlmayr}(1998)}]{Gail1998}
{Gail}, H.-P. \& {Sedlmayr}, E. 1998, Faraday Discussions, 109, 303

\bibitem[{{Gobrecht} {et~al.}(2016){Gobrecht}, {Cherchneff}, {Sarangi},
  {Plane}, \& {Bromley}}]{Gobrecht2016}
{Gobrecht}, D., {Cherchneff}, I., {Sarangi}, A., {Plane}, J.~M.~C., \&
  {Bromley}, S.~T. 2016, \aap, 585, A6

\bibitem[{{Gonz{\'a}lez Delgado} {et~al.}(2003){Gonz{\'a}lez Delgado},
  {Olofsson}, {Kerschbaum}, {Sch{\"o}ier}, {Lindqvist}, \&
  {Groenewegen}}]{GonzalezDelgado2003}
{Gonz{\'a}lez Delgado}, D., {Olofsson}, H., {Kerschbaum}, F., {et~al.} 2003,
  \aap, 411, 123

\bibitem[{{Goumans} \& {Bromley}(2012)}]{Goumans2012}
{Goumans}, T.~P.~M. \& {Bromley}, S.~T. 2012, \mnras, 420, 3344

\bibitem[{{Griffin} {et~al.}(2010){Griffin}, {Abergel}, {Abreu}, {Ade},
  {Andr{\'e}}, {Augueres}, {Babbedge}, {Bae}, {Baillie}, {Baluteau}, {Barlow},
  {Bendo}, {Benielli}, {Bock}, {Bonhomme}, {Brisbin}, {Brockley-Blatt},
  {Caldwell}, {Cara}, {Castro-Rodriguez}, {Cerulli}, {Chanial}, {Chen},
  {Clark}, {Clements}, {Clerc}, {Coker}, {Communal}, {Conversi}, {Cox},
  {Crumb}, {Cunningham}, {Daly}, {Davis}, {de Antoni}, {Delderfield}, {Devin},
  {di Giorgio}, {Didschuns}, {Dohlen}, {Donati}, {Dowell}, {Dowell}, {Duband},
  {Dumaye}, {Emery}, {Ferlet}, {Ferrand}, {Fontignie}, {Fox}, {Franceschini},
  {Frerking}, {Fulton}, {Garcia}, {Gastaud}, {Gear}, {Glenn}, {Goizel},
  {Griffin}, {Grundy}, {Guest}, {Guillemet}, {Hargrave}, {Harwit}, {Hastings},
  {Hatziminaoglou}, {Herman}, {Hinde}, {Hristov}, {Huang}, {Imhof}, {Isaak},
  {Israelsson}, {Ivison}, {Jennings}, {Kiernan}, {King}, {Lange}, {Latter},
  {Laurent}, {Laurent}, {Leeks}, {Lellouch}, {Levenson}, {Li}, {Li},
  {Lilienthal}, {Lim}, {Liu}, {Lu}, {Madden}, {Mainetti}, {Marliani}, {McKay},
  {Mercier}, {Molinari}, {Morris}, {Moseley}, {Mulder}, {Mur}, {Naylor},
  {Nguyen}, {O'Halloran}, {Oliver}, {Olofsson}, {Olofsson}, {Orfei}, {Page},
  {Pain}, {Panuzzo}, {Papageorgiou}, {Parks}, {Parr-Burman}, {Pearce},
  {Pearson}, {P{\'e}rez-Fournon}, {Pinsard}, {Pisano}, {Podosek}, {Pohlen},
  {Polehampton}, {Pouliquen}, {Rigopoulou}, {Rizzo}, {Roseboom}, {Roussel},
  {Rowan-Robinson}, {Rownd}, {Saraceno}, {Sauvage}, {Savage}, {Savini},
  {Sawyer}, {Scharmberg}, {Schmitt}, {Schneider}, {Schulz}, {Schwartz},
  {Shafer}, {Shupe}, {Sibthorpe}, {Sidher}, {Smith}, {Smith}, {Smith},
  {Spencer}, {Stobie}, {Sudiwala}, {Sukhatme}, {Surace}, {Stevens}, {Swinyard},
  {Trichas}, {Tourette}, {Triou}, {Tseng}, {Tucker}, {Turner}, {Vaccari},
  {Valtchanov}, {Vigroux}, {Virique}, {Voellmer}, {Walker}, {Ward}, {Waskett},
  {Weilert}, {Wesson}, {White}, {Whitehouse}, {Wilson}, {Winter}, {Woodcraft},
  {Wright}, {Xu}, {Zavagno}, {Zemcov}, {Zhang}, \& {Zonca}}]{Griffin2010}
{Griffin}, M.~J., {Abergel}, A., {Abreu}, A., {et~al.} 2010, \aap, 518, L3

\bibitem[{{Groenewegen} {et~al.}(2011){Groenewegen}, {Waelkens}, {Barlow},
  {Kerschbaum}, {Garcia-Lario}, {Cernicharo}, {Blommaert}, {Bouwman}, {Cohen},
  {Cox}, {Decin}, {Exter}, {Gear}, {Gomez}, {Hargrave}, {Henning},
  {Hutsem{\'e}kers}, {Ivison}, {Jorissen}, {Krause}, {Ladjal}, {Leeks}, {Lim},
  {Matsuura}, {Naz{\'e}}, {Olofsson}, {Ottensamer}, {Polehampton}, {Posch},
  {Rauw}, {Royer}, {Sibthorpe}, {Swinyard}, {Ueta}, {Vamvatira-Nakou},
  {Vandenbussche}, {van de Steene}, {van Eck}, {van Hoof}, {van Winckel},
  {Verdugo}, \& {Wesson}}]{Groenewegen2011}
{Groenewegen}, M.~A.~T., {Waelkens}, C., {Barlow}, M.~J., {et~al.} 2011, \aap,
  526, A162

\bibitem[{{G{\"u}sten} {et~al.}(2006){G{\"u}sten}, {Nyman}, {Schilke},
  {Menten}, {Cesarsky}, \& {Booth}}]{Gusten2006}
{G{\"u}sten}, R., {Nyman}, L.~{\AA}., {Schilke}, P., {et~al.} 2006, \aap, 454,
  L13

\bibitem[{{Helou} \& {Walker}(1988)}]{1988IRASP.C......0J}
{Helou}, G. \& {Walker}, D.~W., eds. 1988, {Infrared astronomical satellite
  (IRAS) catalogs and atlases. Volume 7: The small scale structure catalog},
  Vol.~7, 1--265

\bibitem[{{Henning} \& {Stognienko}(1996)}]{Henning1996}
{Henning}, T. \& {Stognienko}, R. 1996, \aap, 311, 291

\bibitem[{{H{\"o}fner}(2008)}]{Hofner2008}
{H{\"o}fner}, S. 2008, \aap, 491, L1

\bibitem[{{H{\"o}fner} \& {Andersen}(2007)}]{Hofner2007}
{H{\"o}fner}, S. \& {Andersen}, A.~C. 2007, \aap, 465, L39

\bibitem[{{H{\"o}fner} {et~al.}(2016){H{\"o}fner}, {Bladh}, {Aringer}, \&
  {Ahuja}}]{Hofner2016}
{H{\"o}fner}, S., {Bladh}, S., {Aringer}, B., \& {Ahuja}, R. 2016, \aap, 594,
  A108

\bibitem[{{Justtanont} {et~al.}(2012){Justtanont}, {Khouri}, {Maercker},
  {Alcolea}, {Decin}, {Olofsson}, {Sch{\"o}ier}, {Bujarrabal}, {Marston},
  {Teyssier}, {Cernicharo}, {Dominik}, {de Koter}, {Melnick}, {Menten},
  {Neufeld}, {Planesas}, {Schmidt}, {Szczerba}, \& {Waters}}]{Justtanont2012}
{Justtanont}, K., {Khouri}, T., {Maercker}, M., {et~al.} 2012, \aap, 537, A144

\bibitem[{{Kerschbaum} \& {Olofsson}(1999)}]{Kerschbaum1999}
{Kerschbaum}, F. \& {Olofsson}, H. 1999, \aaps, 138, 299

\bibitem[{{Kessler} {et~al.}(1996){Kessler}, {Steinz}, {Anderegg}, {Clavel},
  {Drechsel}, {Estaria}, {Faelker}, {Riedinger}, {Robson}, {Taylor}, \&
  {Xim{\'e}nez de Ferr{\'a}n}}]{Kessler1996}
{Kessler}, M.~F., {Steinz}, J.~A., {Anderegg}, M.~E., {et~al.} 1996, \aap, 315,
  L27

\bibitem[{{Khouri}(2014)}]{KhouriPhD}
{Khouri}, T. 2014, PhD thesis, Astronomical Institute Anton Pannekoek,
  University of Amsterdam

\bibitem[{{Khouri} {et~al.}(2014){Khouri}, {de Koter}, {Decin}, {Waters},
  {Lombaert}, {Royer}, {Swinyard}, {Barlow}, {Alcolea}, {Blommaert},
  {Bujarrabal}, {Cernicharo}, {Groenewegen}, {Justtanont}, {Kerschbaum},
  {Maercker}, {Marston}, {Matsuura}, {Melnick}, {Menten}, {Olofsson},
  {Planesas}, {Polehampton}, {Posch}, {Schmidt}, {Szczerba}, {Vandenbussche},
  \& {Yates}}]{Khouri2014a}
{Khouri}, T., {de Koter}, A., {Decin}, L., {et~al.} 2014, \aap, 561, A5

\bibitem[{{Khouri} {et~al.}(2016){Khouri}, {Maercker}, {Waters}, {Vlemmings},
  {Kervella}, {de Koter}, {Ginski}, {De Beck}, {Decin}, {Min}, {Dominik},
  {O'Gorman}, {Schmid}, {Lombaert}, \& {Lagadec}}]{Khouri2016}
{Khouri}, T., {Maercker}, M., {Waters}, L.~B.~F.~M., {et~al.} 2016, \aap, 591,
  A70

\bibitem[{{Khouri} {et~al.}(2015){Khouri}, {Waters}, {de Koter}, {Decin},
  {Min}, {de Vries}, {Lombaert}, \& {Cox}}]{Khouri2015}
{Khouri}, T., {Waters}, L.~B.~F.~M., {de Koter}, A., {et~al.} 2015, \aap, 577,
  A114

\bibitem[{{Klein} {et~al.}(2014){Klein}, {Ciechanowicz}, {Leinz}, {Heyminck},
  {G\"usten}, {Kasemann}, {Wunsch}, {Maier}, \& {Sekimoto}}]{Klein2014}
{Klein}, T., {Ciechanowicz}, M., {Leinz}, C., {et~al.} 2014, ITTST, 4, 588

\bibitem[{{Knapp} {et~al.}(2003){Knapp}, {Pourbaix}, {Platais}, \&
  {Jorissen}}]{Knapp2003}
{Knapp}, G.~R., {Pourbaix}, D., {Platais}, I., \& {Jorissen}, A. 2003, \aap,
  403, 993

\bibitem[{{Koike} {et~al.}(1995){Koike}, {Kaito}, {Yamamoto}, {Shibai},
  {Kimura}, \& {Suto}}]{Koike1995}
{Koike}, C., {Kaito}, C., {Yamamoto}, T., {et~al.} 1995, \icarus, 114, 203

\bibitem[{{Kwok}(1975)}]{Kwok1975}
{Kwok}, S. 1975, \apj, 198, 583

\bibitem[{{Li} {et~al.}(2016){Li}, {Millar}, {Heays}, {Walsh}, {van Dishoeck},
  \& {Cherchneff}}]{Li2016}
{Li}, X., {Millar}, T.~J., {Heays}, A.~N., {et~al.} 2016, \aap, 588, A4

\bibitem[{{Lombaert} {et~al.}(2013){Lombaert}, {Decin}, {de Koter},
  {Blommaert}, {Royer}, {De Beck}, {de Vries}, {Khouri}, \&
  {Min}}]{Lombaert2013}
{Lombaert}, R., {Decin}, L., {de Koter}, A., {et~al.} 2013, \aap, 554, A142

\bibitem[{{Lombaert} {et~al.}(2016){Lombaert}, {Decin}, {Royer}, {de Koter},
  {Cox}, {Gonz{\'a}lez-Alfonso}, {Neufeld}, {De Ridder}, {Ag{\'u}ndez},
  {Blommaert}, {Khouri}, {Groenewegen}, {Kerschbaum}, {Cernicharo},
  {Vandenbussche}, \& {Waelkens}}]{Lombaert2016}
{Lombaert}, R., {Decin}, L., {Royer}, P., {et~al.} 2016, \aap, 588, A124

\bibitem[{{Lucas} {et~al.}(1992){Lucas}, {Bujarrabal}, {Guilloteau},
  {Bachiller}, {Baudry}, {Cernicharo}, {Delannoy}, {Forveille}, {Guelin}, \&
  {Radford}}]{Lucas1992}
{Lucas}, R., {Bujarrabal}, V., {Guilloteau}, S., {et~al.} 1992, \aap, 262, 491

\bibitem[{{Maercker} {et~al.}(2016){Maercker}, {Danilovich}, {Olofsson}, {De
  Beck}, {Justtanont}, {Lombaert}, \& {Royer}}]{Maercker2016}
{Maercker}, M., {Danilovich}, T., {Olofsson}, H., {et~al.} 2016, \aap, 591, A44

\bibitem[{{Mamon} {et~al.}(1988){Mamon}, {Glassgold}, \& {Huggins}}]{Mamon1988}
{Mamon}, G.~A., {Glassgold}, A.~E., \& {Huggins}, P.~J. 1988, \apj, 328, 797

\bibitem[{{Mathis} {et~al.}(1977){Mathis}, {Rumpl}, \&
  {Nordsieck}}]{Mathis1977}
{Mathis}, J.~S., {Rumpl}, W., \& {Nordsieck}, K.~H. 1977, \apj, 217, 425

\bibitem[{{Mattsson} {et~al.}(2010){Mattsson}, {Wahlin}, \&
  {H{\"o}fner}}]{Mattsson2010}
{Mattsson}, L., {Wahlin}, R., \& {H{\"o}fner}, S. 2010, \aap, 509, A14

\bibitem[{{Menten} {et~al.}(2010){Menten}, {Bujarrabal}, {Alcolea},
  {Cernicharo}, {Decin}, {Dominik}, {Justtanont}, {de Koter}, {Marston},
  {Melnick}, {Menten}, {Neufeld}, {Olofsson}, {Planesas}, {Pulecka}, {Schmidt},
  {Schoier}, {Szczerba}, {Teyssier}, \& {Waters}}]{Menten2010}
{Menten}, K., {Bujarrabal}, V., {Alcolea}, J., {et~al.} 2010, in COSPAR
  Meeting, Vol.~38, 38th COSPAR Scientific Assembly, 3

\bibitem[{{Min} {et~al.}(2009){Min}, {Dullemond}, {Dominik}, {de Koter}, \&
  {Hovenier}}]{Min2009}
{Min}, M., {Dullemond}, C.~P., {Dominik}, C., {de Koter}, A., \& {Hovenier},
  J.~W. 2009, \aap, 497, 155

\bibitem[{{Min} {et~al.}(2003){Min}, {Hovenier}, \& {de Koter}}]{Min2003}
{Min}, M., {Hovenier}, J.~W., \& {de Koter}, A. 2003, \aap, 404, 35

\bibitem[{{Morris} {et~al.}(1985){Morris}, {Lucas}, \& {Omont}}]{Morris1985}
{Morris}, M., {Lucas}, R., \& {Omont}, A. 1985, \aap, 142, 107

\bibitem[{{Mueller} {et~al.}(2014){Mueller}, {Jellema}, {Olberg}, {Moreno}, \&
  {Teyssier}}]{HIFI}
{Mueller}, M., {Jellema}, W., {Olberg}, M., {Moreno}, R., \& {Teyssier}, D.
  2014, The HIFI Beam: Release 1, Release Note for Astronomers, Tech. Rep.
  HIFI-ICC-RP-2014-001

\bibitem[{{Murakami} {et~al.}(2007){Murakami}, {Baba}, {Barthel}, {Clements},
  {Cohen}, {Doi}, {Enya}, {Figueredo}, {Fujishiro}, {Fujiwara}, {Fujiwara},
  {Garcia-Lario}, {Goto}, {Hasegawa}, {Hibi}, {Hirao}, {Hiromoto}, {Hong},
  {Imai}, {Ishigaki}, {Ishiguro}, {Ishihara}, {Ita}, {Jeong}, {Jeong},
  {Kaneda}, {Kataza}, {Kawada}, {Kawai}, {Kawamura}, {Kessler}, {Kester},
  {Kii}, {Kim}, {Kim}, {Kobayashi}, {Koo}, {Kwon}, {Lee}, {Lorente}, {Makiuti},
  {Matsuhara}, {Matsumoto}, {Matsuo}, {Matsuura}, {M{\"u}ller}, {Murakami},
  {Nagata}, {Nakagawa}, {Naoi}, {Narita}, {Noda}, {Oh}, {Ohnishi}, {Ohyama},
  {Okada}, {Okuda}, {Oliver}, {Onaka}, {Ootsubo}, {Oyabu}, {Pak}, {Park},
  {Pearson}, {Rowan-Robinson}, {Saito}, {Sakon}, {Salama}, {Sato}, {Savage},
  {Serjeant}, {Shibai}, {Shirahata}, {Sohn}, {Suzuki}, {Takagi}, {Takahashi},
  {Tanab{\'e}}, {Takeuchi}, {Takita}, {Thomson}, {Uemizu}, {Ueno}, {Usui},
  {Verdugo}, {Wada}, {Wang}, {Watabe}, {Watarai}, {White}, {Yamamura},
  {Yamauchi}, \& {Yasuda}}]{2007PASJ...59S.369M}
{Murakami}, H., {Baba}, H., {Barthel}, P., {et~al.} 2007, \pasj, 59, S369

\bibitem[{{Mutschke} {et~al.}(1998){Mutschke}, {Begemann}, {Dorschner},
  {Guertler}, {Gustafson}, {Henning}, \& {Stognienko}}]{Mutschke1998}
{Mutschke}, H., {Begemann}, B., {Dorschner}, J., {et~al.} 1998, \aap, 333, 188

\bibitem[{{Norris} {et~al.}(2012){Norris}, {Tuthill}, {Ireland}, {Lacour},
  {Zijlstra}, {Lykou}, {Evans}, {Stewart}, \& {Bedding}}]{Norris2012}
{Norris}, B.~R.~M., {Tuthill}, P.~G., {Ireland}, M.~J., {et~al.} 2012, \nat,
  484, 220

\bibitem[{{Olofsson} {et~al.}(1998){Olofsson}, {Lindqvist}, {Nyman}, \&
  {Winnberg}}]{Olofsson1998}
{Olofsson}, H., {Lindqvist}, M., {Nyman}, L.-A., \& {Winnberg}, A. 1998, \aap,
  329, 1059

\bibitem[{{Poglitsch} {et~al.}(2010){Poglitsch}, {Waelkens}, {Geis},
  {Feuchtgruber}, {Vandenbussche}, {Rodriguez}, {Krause}, {Renotte}, {van
  Hoof}, {Saraceno}, {Cepa}, {Kerschbaum}, {Agn{\`e}se}, {Ali}, {Altieri},
  {Andreani}, {Augueres}, {Balog}, {Barl}, {Bauer}, {Belbachir}, {Benedettini},
  {Billot}, {Boulade}, {Bischof}, {Blommaert}, {Callut}, {Cara}, {Cerulli},
  {Cesarsky}, {Contursi}, {Creten}, {De Meester}, {Doublier}, {Doumayrou},
  {Duband}, {Exter}, {Genzel}, {Gillis}, {Gr{\"o}zinger}, {Henning},
  {Herreros}, {Huygen}, {Inguscio}, {Jakob}, {Jamar}, {Jean}, {de Jong},
  {Katterloher}, {Kiss}, {Klaas}, {Lemke}, {Lutz}, {Madden}, {Marquet},
  {Martignac}, {Mazy}, {Merken}, {Montfort}, {Morbidelli}, {M{\"u}ller},
  {Nielbock}, {Okumura}, {Orfei}, {Ottensamer}, {Pezzuto}, {Popesso},
  {Putzeys}, {Regibo}, {Reveret}, {Royer}, {Sauvage}, {Schreiber}, {Stegmaier},
  {Schmitt}, {Schubert}, {Sturm}, {Thiel}, {Tofani}, {Vavrek}, {Wetzstein},
  {Wieprecht}, \& {Wiezorrek}}]{Poglitsch2010}
{Poglitsch}, A., {Waelkens}, C., {Geis}, N., {et~al.} 2010, \aap, 518, L2

\bibitem[{{Sahai} \& {Bieging}(1993)}]{Sahai1993}
{Sahai}, R. \& {Bieging}, J.~H. 1993, \aj, 105, 595

\bibitem[{{Schoenberg}(1988)}]{Schoenberg1988}
{Schoenberg}, K. 1988, \aap, 195, 198

\bibitem[{{Sch{\"o}ier} {et~al.}(2004){Sch{\"o}ier}, {Olofsson}, {Wong},
  {Lindqvist}, \& {Kerschbaum}}]{Schoier2004}
{Sch{\"o}ier}, F.~L., {Olofsson}, H., {Wong}, T., {Lindqvist}, M., \&
  {Kerschbaum}, F. 2004, \aap, 422, 651

\bibitem[{{Sch{\"o}ier} {et~al.}(2013){Sch{\"o}ier}, {Ramstedt}, {Olofsson},
  {Lindqvist}, {Bieging}, \& {Marvel}}]{Schoier2013}
{Sch{\"o}ier}, F.~L., {Ramstedt}, S., {Olofsson}, H., {et~al.} 2013, \aap, 550,
  A78

\bibitem[{{Skinner} {et~al.}(1999){Skinner}, {Justtanont}, {Tielens}, {Betz},
  {Boreiko}, \& {Baas}}]{Skinner1999}
{Skinner}, C.~J., {Justtanont}, K., {Tielens}, A.~G.~G.~M., {et~al.} 1999,
  \mnras, 302, 293

\bibitem[{{Sloan} {et~al.}(2003){Sloan}, {Kraemer}, {Price}, \&
  {Shipman}}]{Sloan2003}
{Sloan}, G.~C., {Kraemer}, K.~E., {Price}, S.~D., \& {Shipman}, R.~F. 2003,
  \apjs, 147, 379

\bibitem[{{Smith} {et~al.}(2004){Smith}, {Price}, \&
  {Baker}}]{2004ApJS..154..673S}
{Smith}, B.~J., {Price}, S.~D., \& {Baker}, R.~I. 2004, \apjs, 154, 673

\bibitem[{{van der Veen} \& {Breukers}(1989)}]{vanderVeen1989}
{van der Veen}, W.~E.~C.~J. \& {Breukers}, R.~J.~L.~H. 1989, \aap, 213, 133

\bibitem[{{Vassilev} {et~al.}(2008){Vassilev}, {Meledin}, {Lapkin}, {Belitsky},
  {Nystr{\"o}m}, {Henke}, {Pavolotsky}, {Monje}, {Risacher}, {Olberg},
  {Strandberg}, {Sundin}, {Fredrixon}, {Ferm}, {Desmaris}, {Dochev},
  {Pantaleev}, {Bergman}, \& {Olofsson}}]{Vassilev2008}
{Vassilev}, V., {Meledin}, D., {Lapkin}, I., {et~al.} 2008, \aap, 490, 1157

\bibitem[{{Willacy} \& {Millar}(1997)}]{Willacy1997}
{Willacy}, K. \& {Millar}, T.~J. 1997, \aap, 324, 237

\bibitem[{{Woitke}(2006)}]{Woitke2006}
{Woitke}, P. 2006, \aap, 460, L9

\end{thebibliography}

\begin{appendix}

%----------------------------------------------------------------------------------------------------------
\section{Influence of the variable stellar luminosity on the molecular lines}        \label{sect:App:Luminosity}
%----------------------------------------------------------------------------------------------------------

In general, the variability of AGB stars is not taken into account when performing radiative transfer. Their variable luminosity, and therefore variable infrared flux, is hence assumed not to impact the molecular line emission even though infrared pumping is known to affect the excitation of molecular species (see, e.g., \citeauthor{Agundez2006}~\citeyear{Agundez2006}, \citeauthor{Deguchi1984}~\citeyear{Deguchi1984}, \citeauthor{Daniel2012}~\citeyear{Daniel2012}).
As shown by \citet{Cernicharo2014} for the carbon star IRC+10216, integrated line strengths may vary with $L_*$ depending on the molecule and/or the molecular transition. Because the local temperature of the gas could also be modulated by variations in stellar luminosity, both direct radiative pumping and collisional excitation may be affected. 
Accounting for the time-variable stellar luminosity in the radiative-transfer modelling would require the coupling of the calculation of the level populations to the variable stellar flux and the molecular emission from the CSE itself, in which every point sees the stellar flux at a different phase \citep{Cernicharo2014}. This is beyond the scope of this paper. 
We can, however, probe the effect of variable stellar luminosities at constant stellar temperature on the molecular line emission.

\begin{figure}
\centering
\includegraphics[width=.98\columnwidth]{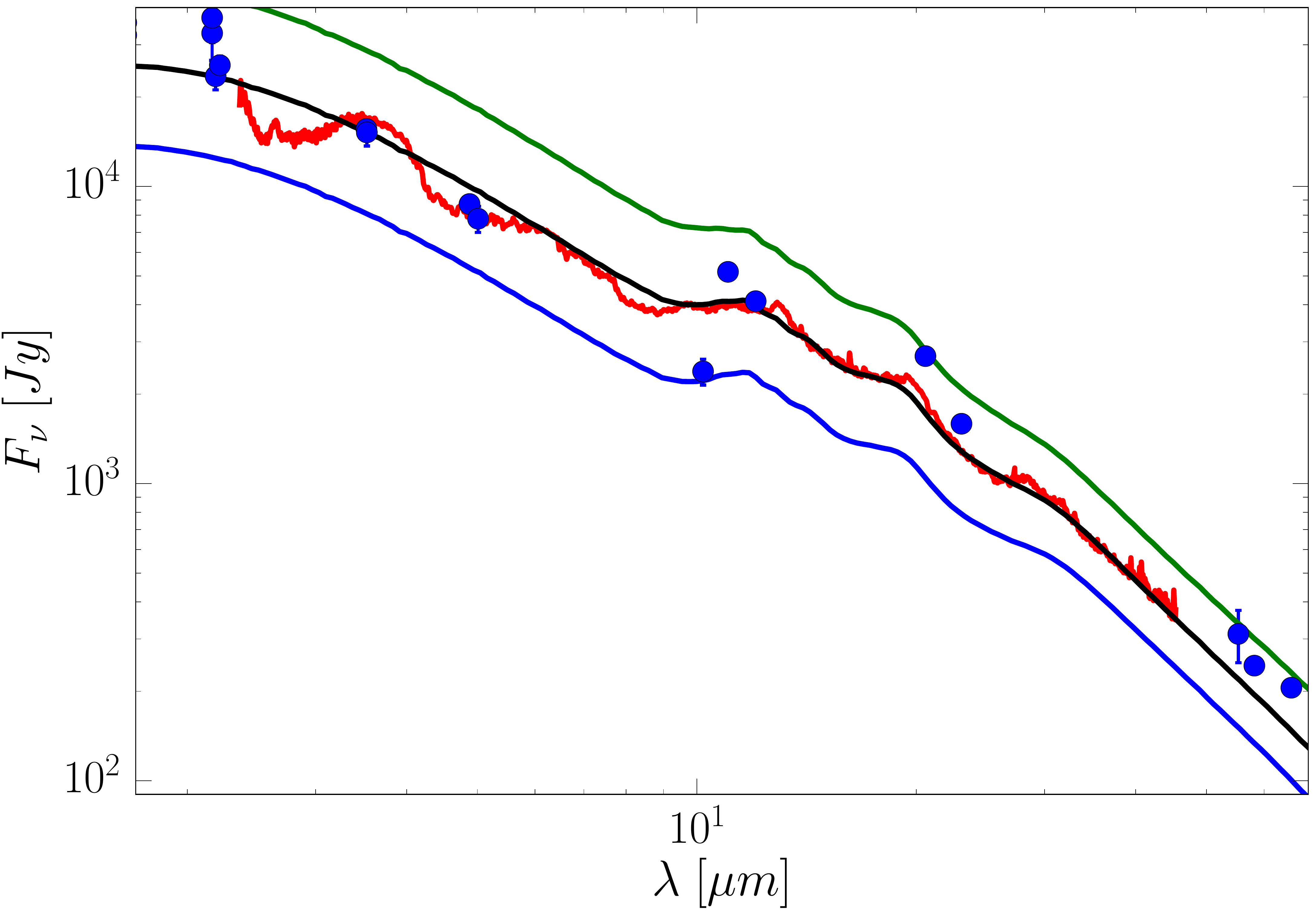}
\caption{Red: ISO-SWS spectrum of R Dor along with photometric data (blue dots). Black: best-fit model with $L_*$= 4500 $L_\odot$ (see Sect. \ref{subsubsect:Results-EnvModel-Dust}). Blue: model with $L_*$ = 2500 $L_\odot$. Green: model with $L_*$= 9000 $L_\odot$.
}
\label{fig:Discussion-L-SED}
\end{figure}

R\,Dor is an SRb-type variable that switches between a primary pulsation mode with a period of 332\,d and a secondary mode of $\sim$175d \citep{Bedding1998}. 
Its light curve is irregular, hence it is not feasible to determine a relation between its phase and luminosity as can be done for Mira variables (see, e.g., \citeauthor{DeBeck2012}~\citeyear{DeBeck2012}). 
Assuming a distance of 59 pc, the stellar luminosity is calculated from the apparent visual magnitude $V$ and its bolometric correction. 
During the pulsation cycle, the overal light distribution may vary as well. This implies that the bolometric correction need not necessarily be constant \citep{Flower1975,vanderVeen1989}.
Putting the latter issue aside, we estimate the ratio of maximum to minimum luminosity $L_{\rm max}/L_{\rm min} = \Delta V / 2.5$ from the variation in apparent visual magnitude $\Delta V$ that can be retrieved from the AAVSO\footnote{https://www.aavso.org/aavso-research-portal} light curve.  Using $\Delta V = 1.5$, we find a luminosity ratio of about 4.  A roughly similar ratio is obtained when using photometric data retrieved from the ViZieR database over the period 1978 $-$ 2012 (see App. \ref{app:PhotometricData}). 
This is shown in Fig. \ref{fig:Discussion-L-SED}. 

\begin{figure}
\centering
\includegraphics[width=.98\columnwidth]{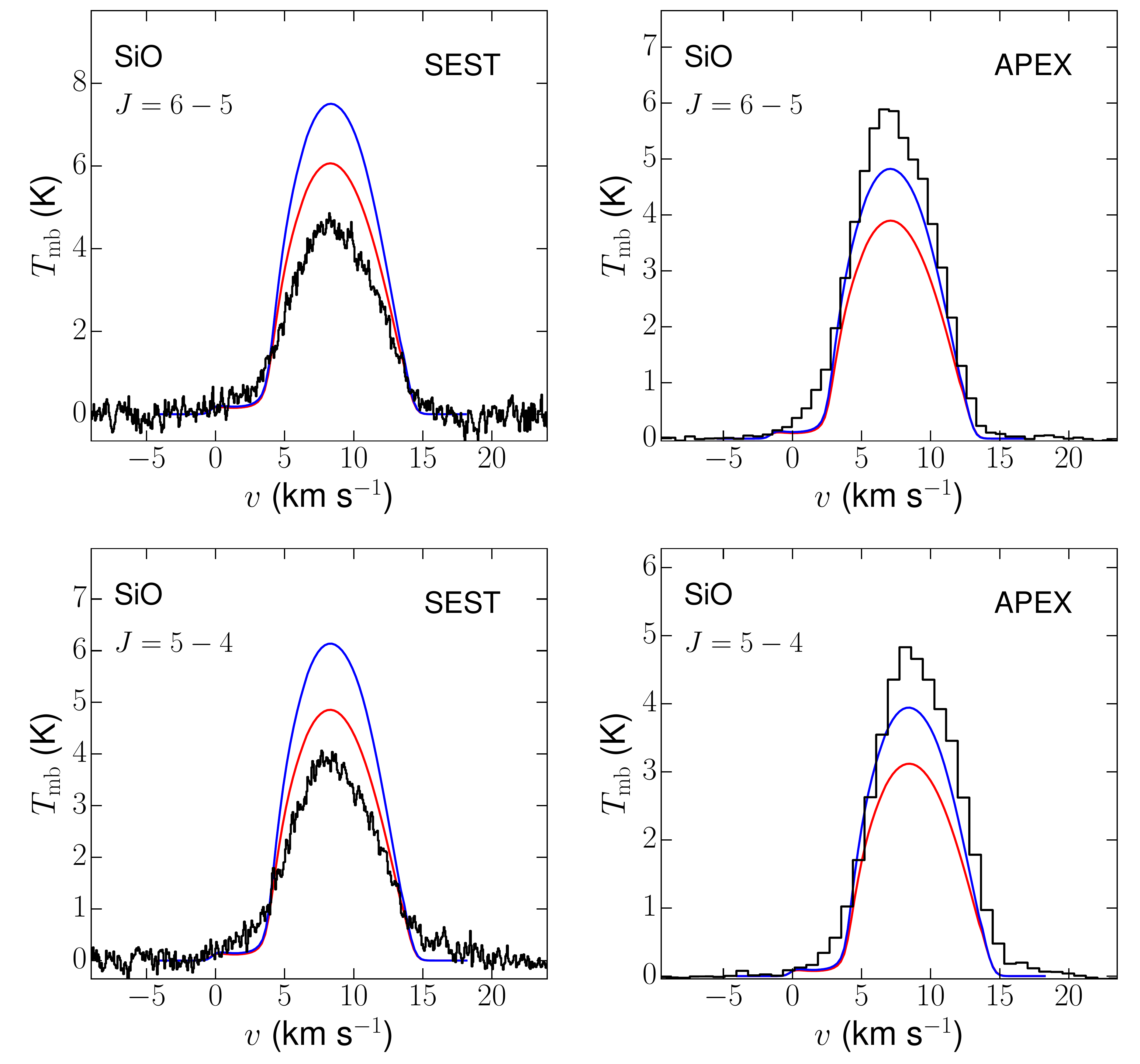}
\caption{Black: resolved line profiles of the SiO $J = 6-5$ and $5-4$ molecular transitions, observed by APEX and SEST in November 2014 and December 1992 respectively.
Blue: predicted line profiles  using $L_* = $ 4500 $L_\odot$ (assumed throughout the modelling) and an abundance profile characterised by $f_0 = 6.0 \times 10^{-5}$ relative to H$_2$ and $e$-folding radius $r_e = 4 \times 10^{13}\ \mathrm{cm}$ with $R_{\rm decl} = 50\ R_*$.
Red: predicted line profiles using $L_* = $ 2500 $L_\odot$ and the same abundance profile.}
\label{fig:Discussion-L-LPs}
\end{figure}

As noted in Sect. \ref{subsect:Results-AbunProf-SiO}, we have excluded the SEST observations of the $J = 6 - 5$ and $5 - 4$ rotational SiO transitions from our analysis. The integrated intensities of the SEST and APEX observations differ by about 20\%, the SEST lines having the lower intensity. We are unable to fit both the APEX and SEST observations simulataneously. The SEST observations date from December 1992 and those of APEX from November 2014. Therefore the differences in the line strengths might be linked to a variable stellar luminosity. 
Indeed, from the AAVSO light curve we derived an approximate difference in apparent visual magnitude of 0.75, which leads to a ratio of the stellar luminosity at the time of these observations of about 2. 
Fig. \ref{fig:Discussion-L-LPs} shows the modelled line profiles for the SiO $J = 6-5$ and $5-4$ SEST and APEX observations for $L_* = 4500 \ L_\odot$ (assumed throughout the modelling) and 2500 $L_\odot$, using the same abundance profile as in Fig. \ref{fig:Results-SiO-LPsGood}. 
Using $L_*$ = 4500 $L_\odot$, the ratio of the integrated line strength of the model to the data lies within 25\% of the integrated line strength for the APEX $J = 6-5$ and $5-4$ transitions, but overestimates the SEST $J = 6-5$ and $5-4$ transitions by an additional 20\%.
For the low luminosity model the APEX transitions are underestimated, and provide a better fit to the SEST transitions by about 10\%.
The discrepancy in integrated line strength of the APEX and SEST $J = 6-5$ and $5-4$ transitions may thus be connected to the variability in the luminosity of R Dor.

\newpage
%%%%%%%%%%%%%%%%%%%%%%%%%%%%%%%%%%%%%%%%%%%%%%%%%%%%%%%
\section{PACS spectra}            \label{app:PACSSpectra}
%%%%%%%%%%%%%%%%%%%%%%%%%%%%%%%%%%%%%%%%%%%%%%%%%%%%%%%

\begin{figure*}[h]
\centering
\includegraphics[angle = 90,width=0.9\textwidth]{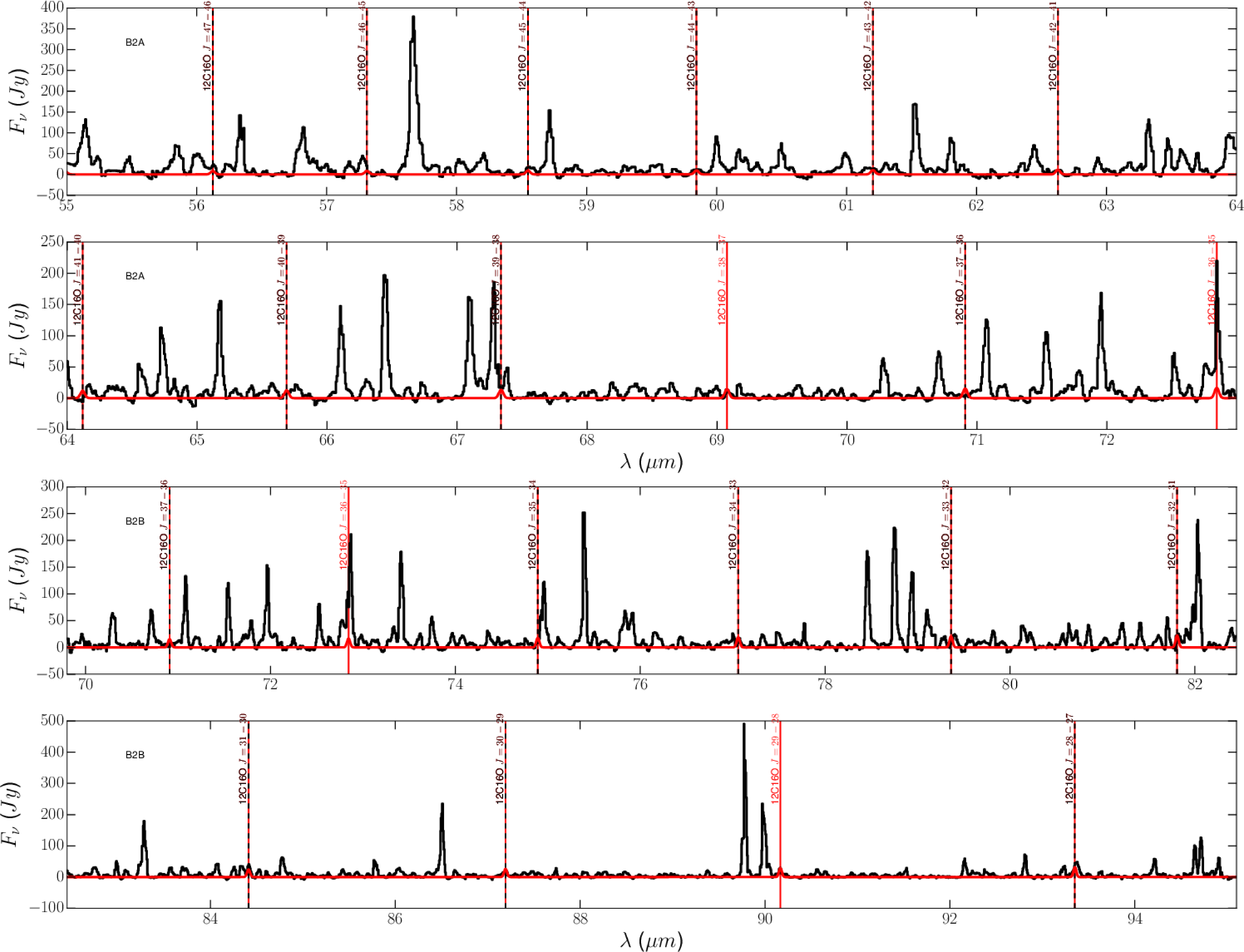}
\caption{Black: continuum-subtracted PACS spectrum of R Dor for the blue bands. Red: modelled spectrum. The PACS band is indicated in the upper left corner of each spectrum. 
The abundance profile used for SiO is characterised by $f_0 = 6.0 \times 10^{-5}$ relative to H$_2$ and $e$-folding radius $r_e = 4 \times 10^{13}$ cm with $R_{\rm decl} = 50\ R_*$. The abundance profile used for HCN is characterised by $f_0$ = $5.5\times 10^{-7}$ with respect to H$_2$  and $r_e = 1.5 \times 10^{15}$ cm. The modelled transitions are pointed out by coloured vertical lines: CO in red, SiO in green, and HCN in blue. Dashed lines indicate modelled transitions that were not detected or contribute to a blended line.
}
\label{fig:App-PACS-B}
\end{figure*}

\begin{figure*}
\centering
\includegraphics[angle = 90,width=0.9\textwidth]{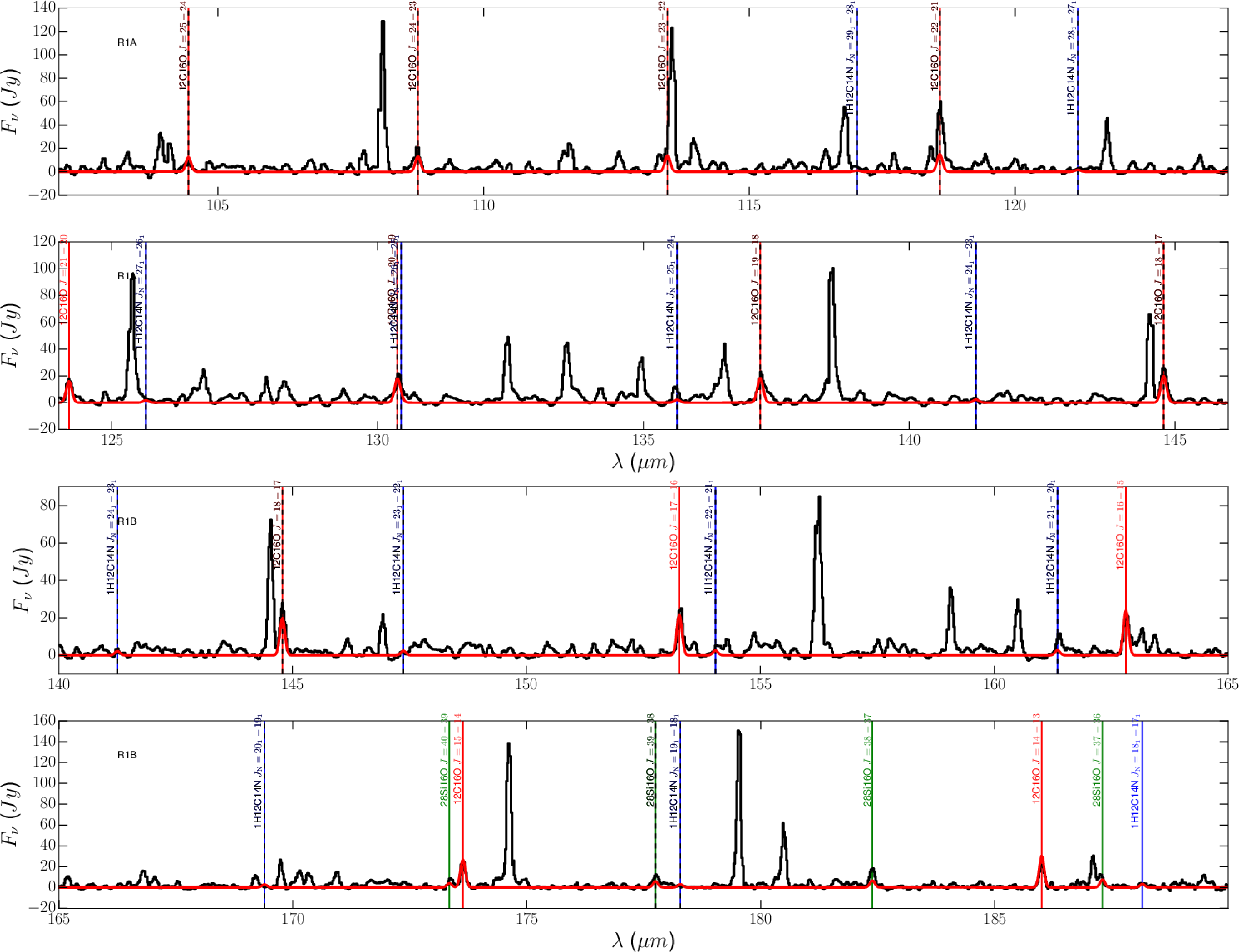}
\caption{Black: continuum-subtracted PACS spectrum of R Dor for the red bands. Red: modelled spectrum. The PACS band is indicated in the upper left corner of each spectrum. 
The abundance profile used for SiO is characterised by $f_0 = 6.0 \times 10^{-5}$ relative to H$_2$ and $e$-folding radius $r_e = 4 \times 10^{13}$ cm with $R_{\rm decl} = 50\ R_*$. The abundance profile used for HCN is characterised by $f_0$ = $5.5\times 10^{-7}$ with respect to H$_2$  and $r_e = 1.5 \times 10^{15}$ cm. The modelled transitions are pointed out by coloured vertical lines: CO in red, SiO in green, and HCN in blue. Dashed lines indicate modelled transitions that were not detected or contribute to a blended line.
}
\label{fig:App-PACS-R}
\end{figure*}

Fig. \ref{fig:App-PACS-B} and \ref{fig:App-PACS-R} show the continuum-subtracted PACS spectrum of R Dor, together with the modelled transitions of CO, SiO, and HCN. 

\newpage
%%%%%%%%%%%%%%%%%%%%%%%%%%%%%%%%%%%%%%%%%%%%%%%%%%%%%%%
\section{Photometric data}            \label{app:PhotometricData}
%%%%%%%%%%%%%%%%%%%%%%%%%%%%%%%%%%%%%%%%%%%%%%%%%%%%%%%
Table \ref{table:Data-phot} lists the photometric data of R Dor used in this work.

\begin{table}[h]
\centering
    \caption{Photometric data of R Dor used in this work. The wavelength, measured flux and its uncertainty, and reference to the measurement are listed. 
    }
    \centering
	\resizebox{0.98\columnwidth}{!}{%
    \begin{tabular}{c c c l}
    \hline \hline 
    \noalign{\smallskip}
    Photomeric          & Flux       & Error         & Reference \\
    band    & [Jy]       & [Jy]             & \\
    \hline
    \noalign{\smallskip}
2MASS K & 3.279e+04 & 6.221e+03 & {\citet{2003yCat.2246....0C}} \\
2MASS K & 3.696e+04 & 6.502e+03 & {\citet{2012yCat.2281....0C}} \\
Johnson K & 2.349e+04 & 2.349e+03 & {\citet{2002yCat.2237....0D}} \\
DIRBE F2.2 & 2.560e+04 & 1.898e+03 & {\citet{2004ApJS..154..673S}} \\
DIRBE F3.5 & 1.561e+04 & 9.548e+02 & {\citet{2004ApJS..154..673S}} \\
Johnson L & 1.520e+04 & 1.520e+03 & {\citet{2002yCat.2237....0D}} \\
DIRBE F4.9  & 8.729e+03 & 2.561e+02 & {\citet{2004ApJS..154..673S}} \\
Johnson M & 7.788e+03 & 7.788e+02 & {\citet{2002yCat.2237....0D}} \\
Johnson N & 2.385e+03 & 2.385e+02 & {\citet{2002yCat.2237....0D}} \\
IRAS F12 & 5.160e+03 & 2.064e+02 & {\citet{1988IRASP.C......0J}} \\
DIRBE F12 & 4.112e+03 & 7.270e+01 & {\citet{2004ApJS..154..673S}} \\
DIRBE F25 & 2.686e+03 & 6.310e+01 & {\citet{2004ApJS..154..673S}} \\
IRAS F25 & 1.590e+03 & 4.770e+01 & {\citet{1988IRASP.C......0J}} \\
DIRBE F60 & 3.120e+02 & 6.230e+01 & {\citet{2004ApJS..154..673S}} \\
IRAS F60 & 2.440e+02 & 1.220e+01 & {\citet{1988IRASP.C......0J}} \\
AKARI N60 & 2.056e+02 & 8.590e+00 & {\citet{2007PASJ...59S.369M}} \\

    \hline
    \end{tabular}%
    }

    \label{table:Data-phot}
\end{table}

\end{appendix}

\end{document}